# Morphology of Thin Films of Gold Grown on CaF$_2$ (100) and (111)

*Georg Albert, Marco Maccarini, and Michael Himmelhaus\**

Angewandte Physikalische Chemie, Universität Heidelberg, Im Neuenheimer Feld 253, 69120 Heidelberg

**Abstract.**

Thin films of gold were deposited on single crystalline CaF$_2$ substrates of (100) and (111) orientation by thermal evaporation. The resulting film structure was analyzed by means of atomic force microscopy in dependence of substrate temperature and deposition thickness. A marked substrate dependency was found. On CaF$_2$ (100), extended and flat plateaus with diameters of several hundreds of nanometers form in a network-like arrangement for deposition thicknesses between 30 – 45 nm. On CaF$_2$ (111), closed films of small average roughness develop at film thicknesses between 70 – 90 nm. These films do not exhibit, however, the extended plateaus present in the former case. On basis of the rule of Brück and Engel, epitaxial growth of Au (100) crystallites on both surfaces is suggested. For both substrates, the epitaxial temperature was determined to $T_e = 320º$ C.

PACS codes, keywords

\*Corresponding author: email: il7@ix.urz.uni-heidelberg.de

**1.     Introduction.**

Numerous efforts in nanotechnology were recently devoted to control surfaces on decreasing scale. At the same time, the need for a wider lateral extent of the overall structures fabricated became increasingly important. This poses demands for substrates with well-defined structure over many orders of magnitude, ranging from the nanometer to macroscopic scale. One common route of functionalization of such surfaces, particularly in nanolithography and bionanotechnology,[1] is the utilization self-assembled monolayers (SAM), such as alkyl silanes on metal oxides and alkanethiolates on the coinage metals,[2] which allow spontaneous formation of ultrathin organic films with well-defined properties. In particular alkanethiolates on the coinage metals, such as gold and silver, have proven to form highly ordered monolayers of homogeneous structure, since the sulfur-metal bond provides an enthalpic gain of the order of a covalent bond, but does not exhibit the strong local confinement of the latter. This allows thiolates to rearrange on the surface after adsorption, thereby forming domains of highly ordered molecules. The size of such domains is mainly limited by the limited size of the granules of the gold film, since - for costs reasons - mainly evaporated and thus polycrystalline films are utilized. The size of the granules is basically governed by the growth mode of the metal on the respective substrate and should be largest for epitaxial growth. So far, however, gold has been found to exhibit epitaxial growth only on few substrates, such as mica,[3] molybenide,[4] magnesium oxide,[5] germanium,[5] and alkali halides.[4,5] On mica, (111) oriented crystallites of large extension can be obtained by a number of techniques, such as flame annealing[3] or template stripping.[6] The resultant gold coated substrates have become popular notably in applications where a certain waviness of the substrate on mesoscopic and macroscopic scale can be tolerated. This is in particular true for scanning probe techniques, such as STM and AFM,[7] and for electron spectroscopies, like XPS and NEXAFS.[8] Optical techniques, however, rely on proper reflection or transmission of the probe beams, without affecting polarization and/or reflectivity. Therefore, optical analysis requires flat substrates also on macroscopic scale. In addition, it is often desirable to deal with a substrate transparent to the probe beam, either to improve access to the surface layer, for example when utilizing total attenuated reflection (ATR), or to minimize damage of the surface due to absorption, which is an issue in nonlinear optical techniques, such as surface second harmonic generation (SHG) and sum frequency generation (SFG).

In this article we explore the potential of single crystalline calcium fluoride ($CaF_2$) with (100) and (111) orientation as a substrate for the growth of thin gold films in order to obtain laterally extended gold crystallites on microscopic scale in combination with optical flatness of the deposited film in the mesoscopic and macroscopic regime. Accordingly, optical studies of highly ordered, and potentially nanostructured ultrathin organic films, such as differently functionalized thiolate SAMs or surface-grafted (bio-)polymers, will become feasible. We use $CaF_2$ due to its high transmission range from the UV to the MIR region,[9] which recently has attracted increasing attention in linear and nonlinear[10] optical spectroscopy. Further, $Au/CaF_2/Si(111)$ ternary systems showed the formation of flat gold islands,[11,12] while a recent comparative study[13] on the formation of thin gold films on mica, glass, silicon, and polycrystalline $CaF_2$ substrates gave first evidence of epitaxial growth on $CaF_2$.

In the following AFM study, the dependence of film morphology on temperature and nominal gold deposition thickness is analysed. A number of different growth stages will be presented, and conditions for the formation of flat, closed, and epitaxial gold films will be given.

2. **Experimental.**

**Chemicals.** Analytical grade isopropanol (99.9%) was purchased from Sigma-Aldrich and used as received. Gold of (99.99%) purity was obtained from Allgemeine Gold- und Silberscheideanstalt, Pforzheim, Germany.

**Substrates.** Single crystalline $CaF_2$ substrates with (111) and (100) orientation and a size of 10x10x0.5 $mm^3$ were purchased from Crystec GmbH, Berlin, Germany. AFM measurements on both types of surfaces confirmed an rms roughness of about 2 Å on a scale of 1x1 $\mu m^2$, which is better than the specification of < 3.5 Å (typically).

**Cleaning.** Before evaporation, the substrates were sonicated in analytical grade isopropanol for 15 minutes. Just before transferring the samples into the evaporator, samples were blown dry with pure nitrogen. We tried different kinds of wet cleaning procedures, such as alternating sonication in water and isopropanol. However, it seems that water affects the $CaF_2$ surface, leading to less reproducible results in the subsequent evaporation step.

After the base pressure of $5 \times 10^{-8}$ mbar was reached, the substrates were thermally cleaned by keeping them at 350º C for 4-5 hours.

**Evaporation.** After termination of the thermal cleaning of the substrates under UHV conditions, the substrates were slowly brought to the temperature chosen for evaporation. Then, a thin film of gold of chosen thickness was evaporated onto the substrates. After evaporation, the substrates were kept at the given temperature for another 3 hours. Then, the samples were allowed to cool down to room temperature. The chamber of the evaporator was vented with pure Argon. After removal from the evaporator, the samples were also stored under pure Argon.

**Instrumentation.** The evaporator used is home-built and operates at a base pressure of $5 \ 10^{-7}$ mbar by means of a Leybold rotary pump. Tungsten boats were used for evaporation of the gold.

AFM measurements were performed with a Park Scientific (PSI) Autoprobe-GP. The images were taken in contact mode with a cantilever of silicon nitrite from PSI (tip radius 10nm, spring constant 0.2 N/m). Images were processed and analysed with the built in software ProScan version 1.51b (copyright Park Scientific Instruments). Images of larger area

($3\times3$ μm$^2$) were processed by second order flattening. Smaller images were flattened with smaller order (zero or first order).

## 3. Results.

The dependence of film growth on the temperature during the evaporation process was studied at a constant gold deposition thickness of 30 nm. This value was chosen with reference to the deposition of gold films on silicon or glass, which are continuous and flat at a thickness of some tens of nanometers.[14] Figures 1 and 2 display AFM images for CaF$_2$ (100) and (111) surfaces, respectively, and deposition temperatures ranging from 220° to 420° C. For each condition, we present two AFM images, one survey of an area of $3\times3$ μm$^2$ and one detail scan of $100\times100$ nm$^2$ to elucidate the structure of the films on both, mesoscopic and microscopic scale in terms of roughness and plateau formation. The rms values of roughness are summarised in Table 1. Further, a descriptive identifier is given to the different morphologies for their classification following the assignments used in the literature whenever possible.

At 220° C, irregularly arranged crystallites forms on both substrates, suggesting a polycrystalline morphology referred in the following as "small hill" structure. Roughness is lower on (111) oriented substrates, on mesoscopic as well as microscopic scale. Note that roughness of the native CaF$_2$ surfaces after polishing is typically below 3.5 Å on mesoscopic scale, which is almost negligible with respect to the values found after evaporation of the gold. At 280° C, network-like structures ("network") form. Interestingly, the voids on the (100) surface are much larger than the holes found in the film deposited on CaF$_2$(111) ("holes"). The large voids on the (100) surface cause high rms roughness on mesoscopic scale. However, on microscopic scale roughness is in the same range as at 220º C. Again, the (111) surface exhibits the smoother films. At 320º C, the Au films on both surfaces still show the same type of network structure. While on CaF$_2$(111) microscopic roughness is slightly increased, the (100) surface exhibits a minimum value at this temperature. This indicates that large and smooth terraces form. At higher temperature (420º C), both surfaces show an island-type of growth. Here, in particular on $CaF_2$(111), flat and predominantly rounded islands of almost homogeneous height develop. The flat top structure of these islands - reminding of mesas statistically arranged in a plane - suggests a well-defined crystal structure. On the (100) surface, the islands are larger and exhibit channel-like shapes. Despite this difference, they show reasonable flatness, suggesting a similar quality in crystal structure as compared to the crystals formed on the (111) surface.

Since we are interested in the formation of thin continuous films of minimum roughness, we have chosen the deposition temperature of 320º C to explore the dependency of film morphology on thickness, where both surfaces exhibit a roughness close to the respective minimum value. Accordingly, the latter was varied from 5 to 300 nm. Above 90 nm deposition thickness, however, the quality of the deposited films in terms of roughness and plateau formation drops significantly. Therefore, we focus on the region between 5 and 90 nm. Figures 3 and 4 display the AFM images of the resulting gold films on $CaF_2$(100) and (111), respectively. The corresponding structure identifiers and values for roughness on mesoscopic and microscopic scale are given in Table 2.

After deposition of 5 nm of gold at 320º C, "small hill" structures appear on both surfaces similar to those found for deposition of 30 nm of Au at 220º C. However the roughness, in particular on microscopic scale, is significantly higher than that of the former case. After deposition of 10 nm of gold, the crystallite size increases. While the gold film on $CaF_2$(100) exhibits round-shaped crystallites, on the (111) surface it is irregularly shaped, thereby forming a network- or channel-like structure. Mesoscopic roughness increases on both surfaces, whereas microscopic roughness decreases, indicating the presence of flat terraces. As the thickness of the deposited film increases (20 – 45 Å), interconnected films form on both surfaces. In analogy with the study of the temperature dependency, the voids in the films deposited on the $CaF_2(100)$ are much larger than the holes present on the (111) surface. Therefore, we refer to the latter structure as "holes". In this regime, roughness on mesoscopic scale is basically governed by the presence of voids and holes on the two different substrates. On microscopic scale, roughness is decreasing. On $CaF_2(100)$ it exhibits a minimum for a deposition thickness of 30 nm. In the case of the (111) surface, the minimum roughness is found at higher deposition thicknesses.

With increasing gold deposition, the voids and holes in the films close. On the (100) surface, a "rolling hill" morphology can be observed at deposition thicknesses of 70 and 90 nm. In contrast to mesoscopic roughness, which decreases in this stage due to the lack of larger voids, microscopic roughness becomes higher due to the decreasing number of flat terraces. Gold films deposited on $CaF_2(111)$ are substantially hole-free at deposition thicknesses of 70 and 90 nm, however, a drastic change of film morphology is observed. Although the films become very flat with a roughness of about 5 Å on mesoscopic and down to 0.23 Å on microscopic scale, they show many irregularly shaped steps and terraces, thus comprising meander-like morphology.

For illustration of the structural changes induced in the films beyond a thickness of 45 nm, Fig. 5 displays 3D profiles of (100) and (111) surfaces after deposition of 45 and 90 nm gold, respectively, at 320º C. On $CaF_2(100)$, large and flat terraces develop after deposition of 45 nm of gold, despite of the presence of large voids. In the 90 nm film the voids vanish, and the formerly flat terraces tilt with respect to the surface normal, thus increasing surface roughness on microscopic scale. On $CaF_2(111)$ in contrast, smaller terraces form after deposition of 45 nm of gold, which seem to be less flat and comprise an irregular microstructure. At 90 nm, these irregularities have developed. Despite of the overall smoothness of the film, which is indicated by the low RMS values of 5.4 Å on mesoscopic and 0.23 Å on microscopic scale, no regular and flat plateaus are discernible. The latter observation suggests a change of the growth mode somewhere between 45 and 90 nm.

If film thickness is further increased beyond 90 nm, rough films of irregular morphology form. As an example, Fig. 6 displays a 300 nm thick gold film deposited on $CaF_2$(100). While still some terraces are observable, lots of grooves and faults develop in-between, thus randomizing the overall structure. We failed in providing high quality AFM images of 300 nm films grown on $CaF_2$(111) due to a worse film morphology, in agreement with the less regular structure of the films on the (111) surface found at lower deposition thickness.

## 4.    Discussion.

The various structures found for thermal evaporation of gold onto $CaF_2$ (111) and (100) surfaces are in general agreement with the growth process of thin metal films as discussed by Pashley et al.[15] The successive stages of nucleation, growth, and coalescence of islands during the formation of ultrathin continuous metal films were found in numerous studies on related systems, such as gold or silver on glass or mica.[13,16] In particular the regression of the evolution of film formation with increasing deposition temperature, i.e., closed films at low temperature and islands at high deposition temperature, confirms earlier findings.[16]

Besides this general agreement, our study reveals a number of specific properties of the studied system, in particular with respect to the dependency of film morphology and structure on substrate orientation. Of course, these differences could simply arise due to differences in interfacial tension, i.e., wetting, between gold and the two different $CaF_2$ surfaces. However, both are low index surfaces with a difference in lateral packing density of only 15%[17]. The surface tension of liquid $CaF_2$ amounts to approx. 290 mN/m.[18] Gold, however, is known to have a high surface tension of approx. 2000 mN/m in the solid and 1000 mN/m in the liquid state,[9] so that the difference in the two substrate orientations should be negligible with respect to their interaction with the metal. Further, in case of a pure wetting effect, the structure of the gold crystallites should not depend on that of the underlying substrate, neither laterally or vertically. Accordingly, neither any dependence of film structure on film thickness nor the formation of preferentially oriented crystallites should be observable. However, in particular the (100) surfaces exhibit the formation of non-isotropic crystallites, with – at least in some cases – a mesa-like shape (*cf.* Fig. 5a). Such a flat topology suggests a well-defined crystal structure of the crystallites, and thus their epitaxial growth.

On $CaF_2$(100) the gold crystallites are generally larger, so that extended terraces can form. These terraces are surrounded by voids in the film of similar dimension. On $CaF_2$(111), the crystallites formed are smaller and accordingly, also the voids found in the films are small. There are two explanations for this difference. First of all, the number of nucleation sites on the (111) surface could be much larger than in the case of the (100) surface. Secondly, the difference could be caused by specific properties of the respective substrate-adsorbate system, such as differences in surface diffusion or lattice mismatch. For ideal (111) and (100) surfaces, the number of nucleation sites should not defer too much due to the similar packing density[17]. In the case of real surfaces, defects such as dislocations, step edges or point defects, can provide nucleation sites for adsorbed metal atoms. However, the (111) surface is the predominant cleaving surface of $CaF_2$, thus suggesting that the number of defects should be smaller as compared to the (100) surface. Therefore, if surface defects would dominate crystal growth, the (100) surface should exhibit the smaller crystal size. However, this is not the case.

Therefore, we assume that the structural differences found in the films deposited on $CaF_2$ (100) and (111) are caused by specific properties of the respective substrate-adsorbate system. A different surface diffusion of the adsorbed gold atoms on $CaF_2$(100) and (111) might be responsible for the different size of the crystallites, however, can neither explain the different morphologies found on the two surfaces, nor the dependency of film structure on thickness.

According to the rule of Brügg and Engel for epitaxy of metals on alkali halides,[5] those crystal orientations of the deposited metal are preferred during initial growth, which minimize the metal-halide distance at the interface. In our case, this empirical rule leads to the following conclusions.

Both, gold and $CaF_2$, comprise fcc structure. $CaF_2$ is member of the so-called fluorite lattice type, which can be viewed as being composed of a fcc lattice of $Ca^{2+}$ ions, in which the $F^-$ ions fill the tetrahedral voids. The lattice constant amounts to $a_{CaF2}$=5.4623 Å. The lattice constant of gold is $a_{Au}$ =4.078 Å. Therefore, growth of the otherwise preferred Au(111) orientation on $CaF_2$(111) can be excluded due to significant lattice mismatch, which is unlikely according to the rule of Brück and Engel. Here, we assume that gold crystallizes preferably in a way that it exposes low index planes to minimize its surface tension. Therefore, that plane with the lowest index that is commensurable with the substrate lattice, should be the preferred one in epitaxial growth. However, not only on $CaF_2$(111), but also on the (100) surface, which comprises the same $Ca^{2+}$-$Ca^{2+}$ spacing in a square arrangement instead of a hexagonal lattice, no reasonable orientation of the Au(111) plane can be found such that it matches the substrate. From the next lowest gold crystal planes, the (100) plane shows reasonable fit with the $CaF_2$(100) when rotated by 45° with respect to the substrate lattice, as illustrated in Fig. 7a. The misfit is below 5% in this case and the gold lattice should experience a slight compression to match with the $CaF_2$ lattice, thereby revealing that the tilted terraces, which occur at high deposition thickness on $CaF_2$(100) (*cf.* Fig. 5b), may be caused by stress release.

The Au(100) plane matches also the $CaF_2$(111) plane, however, only in one direction with a mismatch below 5% (fig. 7b). In the direction of the second lattice vector, it does not fit, but is commensurable after 3 repeats. This worse situation might explain the growth of smaller crystallites as well as the formation of the meander-like structure at higher deposition thickness. Altogether, according to the rule of Brück and Engel, the differences in the lattice mismatch between gold and $CaF_2$ provide an explanation for the different morphologies, the thin gold films adopt on the two different $CaF_2$ surfaces.

**5.     Conclusions.**

Our study of the deposition of thin gold films on $CaF_2$ single crystals with (100) and (111) orientation reveals specific substrate dependent effects, such as the occurence of flat and extended plateaus of several hundreds of nanometers for gold deposition on the (100) surface and the formation of irregular, however, surprisingly smooth films on $CaF_2(111)$. Both effects show a strong dependence on both, substrate temperature during deposition and thickness of the deposited film. The onset of epitaxial growth was found to be around 320º C. However, one should bear in mind, that the epitaxial temperature strongly depends on the experimental conditions,[5] so that it cannot be generalized. For the dependency on film thickness, the best results in terms of plateau flatness and average roughness were observed between 30 - 45 nm for the (100) and between 70 - 90 nm for the (111) surface. For higher thickness, the evolution of an irregular morphology was observed on both substrates.

While our findings give new insight into the general importance of substrate orientation for the growth of thin metal films on earth alkali halides, they also point out the potential as well as the limitations of the system under study for applications in nanotechnology. Our motivation for this work was to find conditions for the fabrication of ultraflat gold films of large lateral extension, which allow for the formation of highly ordered SAM with a minimum defect density, e.g., at grain boundaries. According to our results, this is achieved, e.g., with the 45 nm films deposited on $CaF_2$ (*cf.* Fig. 5a). Unfortunately, however, these films still show large voids in-between the crystallites, thus restricting potential applications to those, where a potential influence of the bare $CaF_2$ surface area can be tolerated. In constrast, the films deposited on the (111) surface are smooth and closed after deposition of 70 – 90 nm of gold, however, lack the formation of extended plateaus. Therefore, the choice of substrate and resulting film morphology depends strongly on the individual application.

In this AFM study we were not able to identify the orientation of the epitaxially grown crystallites unambiguously. However, application of Brück and Engel's rule suggests that on both substrates (100) oriented gold crystallites prevail. If this conclusion, which needs be confirmed in future studies, proves to be correct, the systems under study would allow an elegant way for the fabrication of thin gold films with an orientation so far accessible only by use of gold single crystals. Thereby, new opportunities for the study of SAM formation on the coinage metals might become feasible.

**Acknowledgment**

This work was funded by the Deutsche Forschungsgemeinschaft under grant no. Hi693/2-1 and Hi693/2-2.

Table 1

| 100 | | | | 111 | | | |
|---|---|---|---|---|---|---|---|
| Temperature | Morphology | RMS 1 | RMS 2 | Temperature | Morphology | RMS 1 | RMS 2 |
| 220 | Small hills | 25.5 | 1.79 | 220 | Small hills | 15 | 0.63 |
| 280 | Network | 267 | 1.72 | 280 | Holes | 37.4 | 0.57 |
| 320 | Network | 368 | 0.2 | 320 | Holes | 50.2 | 1.17 |
| 420 | Channels/islands | 300 | 0.31 | 420 | Islands | 240 | 1.54 |
| | | | | | | | |

Table 2

| 100 | | | | 111 | | | |
|---|---|---|---|---|---|---|---|
| Thickness | Morphology | RMS 1 | RMS 2 | Thickness | Morphology | RMS 1 | RMS 2 |
| 5 | Small hills | 16.5 | 7.3 | 5 | Small hills | 35.1 | 11.4 |
| 10 | Hills | 123 | 1.17 | 10 | Network/channel | 93.7 | 1.7 |
| 20 | Network | 107 | 1.17 | 20 | Holes/channels | 5.14 | - |
| 30 | Network | 368 | 0.2 | 30 | Holes | 50.2 | 1.17 |
| 45 | Network | 265 | 0.44 | 45 | Holes | 58.2 | 0.8 |
| 70 | Rolling Hills | 28 | 0.75 | 70 | Meander | 4.91 | 0.65 |
| 90 | Rolling Hills | 54.1 | 0.67 | 90 | Meander | 5.44 | 0.23 |
| | | | | | | | |

**Figures Captions**

Figure 1. AFM images of 30 nm thick Au films prepared on $CaF_2$ (100) at four different temperatures: a) 220 °C; b) 280 °C; c) 320 °C; d) 420 °C. The right and left images are scans taken on a surface area of 3x3 $\mu m^2$ and 100x100 $nm^2$, respectively.

Figure 2. AFM images of 30 nm thick Au films prepared on $CaF_2$ (111) at four different temperatures: a) 220 °C; b) 280 °C; c) 320 °C; d) 420 °C. The right and left images are scans taken on a surface area of 3x3 $\mu m^2$ and 100x100 $nm^2$, respectively.

Figure3. AFM images of Au films of various thicknesses deposited on $CaF_2$ (100) at a temperature of 320°C: a) 5 nm; b) 10 nm; c) 20 nm; d) 30 nm; e) 45 nm; f) 70nm; g) 90 nm. The right and left images are scans taken on a surface area of 3x3 $\mu m^2$ and 100x100 $nm^2$, respectively.

Figure4. AFM images of Au films of various thicknesses deposited on $CaF_2$ (111) at a temperature of 320°C: a) 5 nm; b) 10 nm; c) 20 nm; d) 30 nm; e) 45 nm; f) 70nm; g) 90 nm. The right and left images are scans taken on a surface area of 3x3 $\mu m^2$ and 100x100 $nm^2$, respectively.

Figure 5. 3D AFM images of Au films deposited on $CaF_2$ substrates at a temperature of 320 °C: a) 45 nm Au on $CaF_2$ (100); b) 90 nm Au on $CaF_2$ (100); c) 45 nm Au on $CaF_2$ (111); d) 90 nm Au on $CaF_2$ (100).

Figure 6. 3D AFM image of a 300 nm thick Au film deposited on $CaF_2$ (100) at 300°C.

Figure 7. Suggested orientation of the Au(100) surface (solid circles) with respect to the underlying CaF$_2$ lattice. (a) Au(100) on CaF$_2$ (100) shows best match according to the rule of Brück and Engel if rotated by 45° with respect to the substrate lattice; (b) Au(111) on CaF$_2$ (111) yields close match with the substrate lattice only in direction of one lattice vector.

**Tables Captions**

Table 1. Rms roughness measured on 30 nm thick Au films on CaF$_2$ (100) and CaF$_2$ (111) at various temperatures. RMS1 and RMS2 indicate rms values measured on a surface area of 3x3 μm$^2$ and 100x100 nm$^2$, respectively.

Table 2. Rms roughness of Au films on CaF$_2$ (100) and CaF$_2$ (111) of various thicknesses prepared at temperature of 320°C. RMS1 and RMS2 indicate rms values measured on a surface area of 3x3 μm$^2$ and 100x100 nm$^2$, respectively.

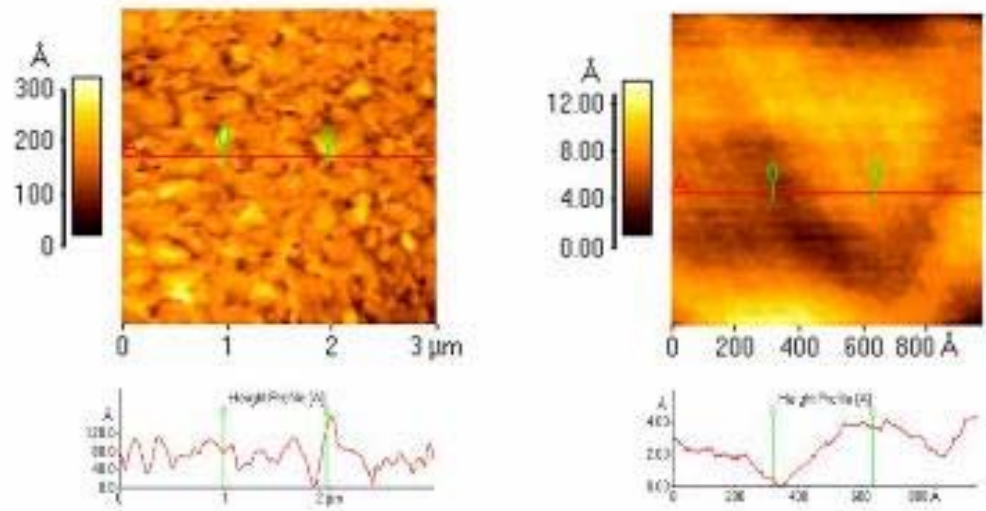

Figure 1 a

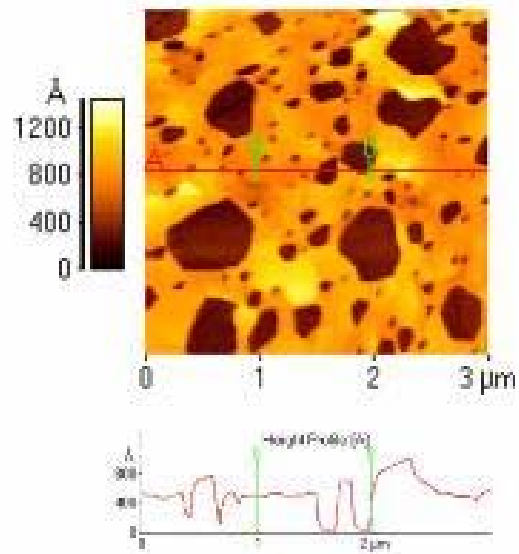 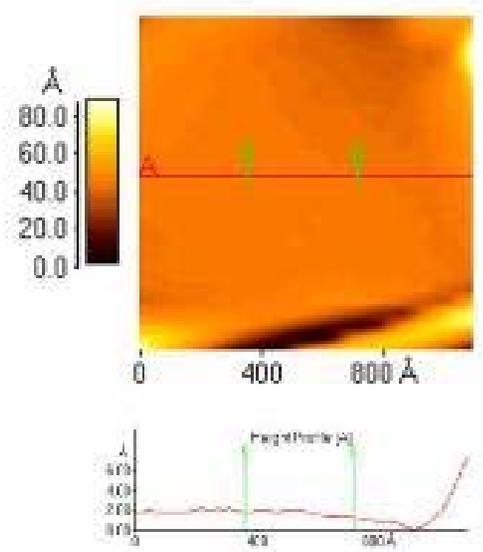

Figure 1 b

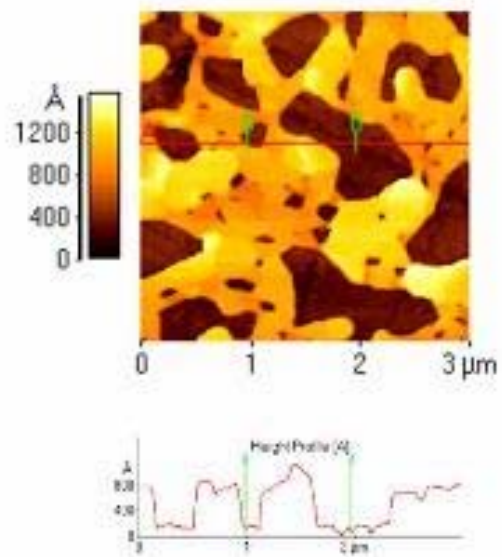 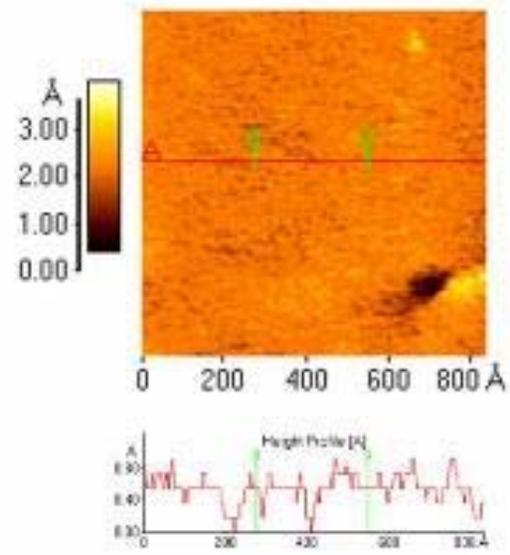

Figure 1 c

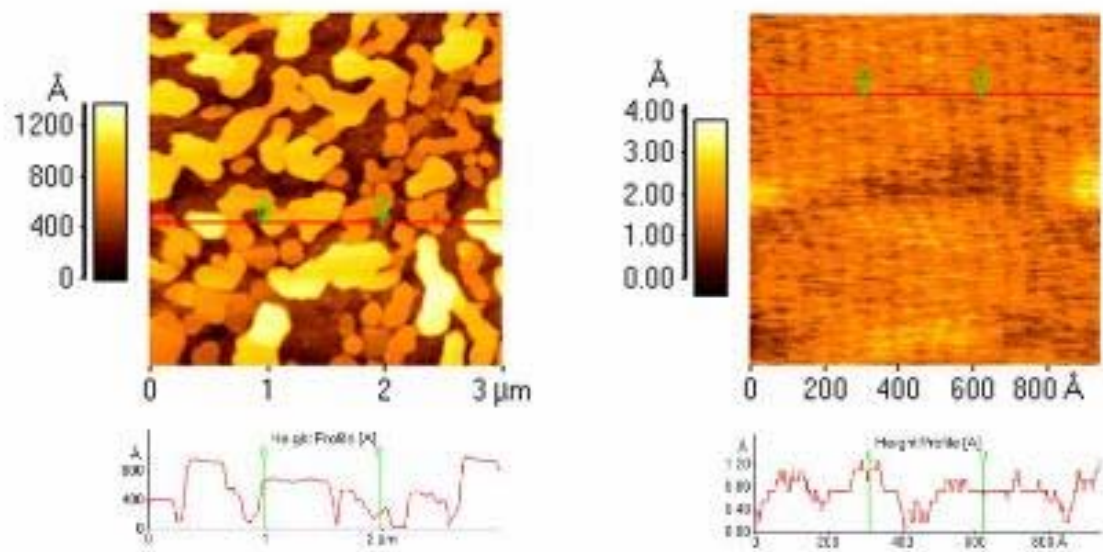

Figure 1 d

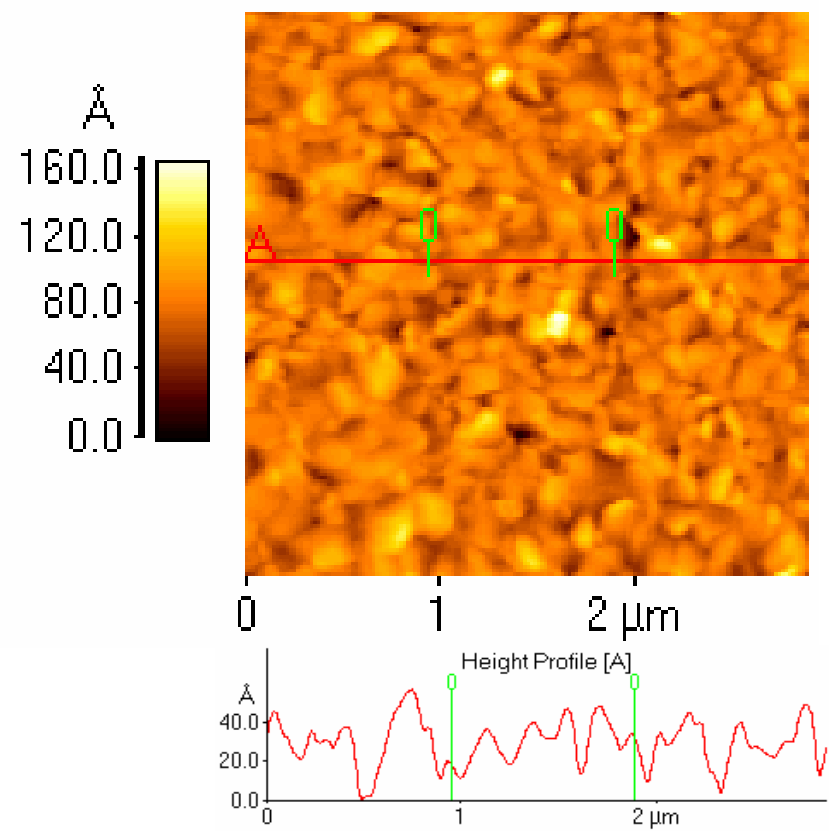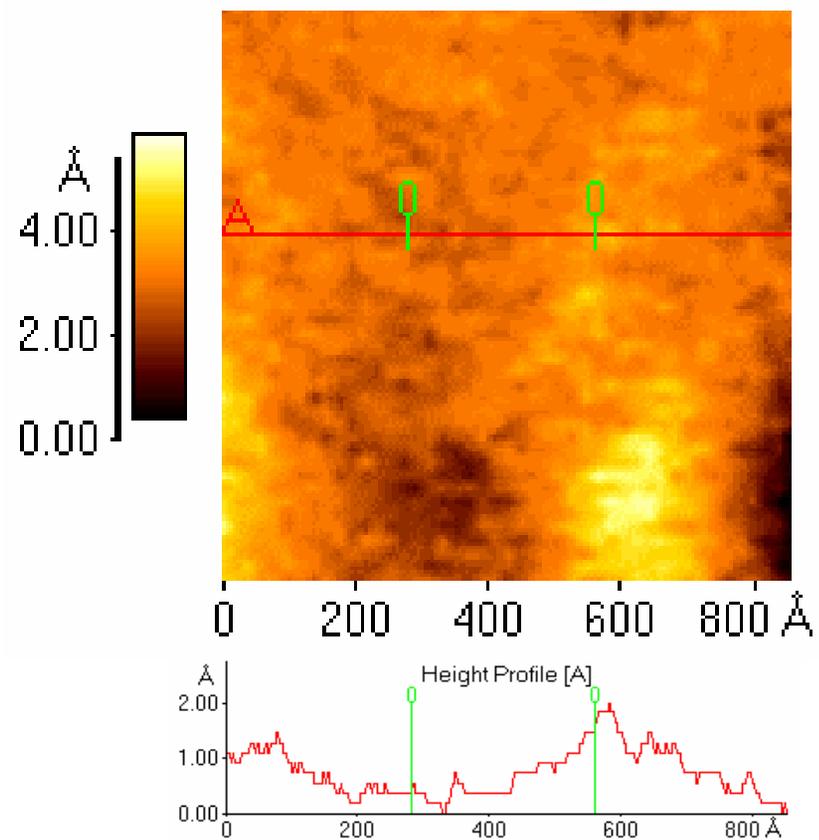

Figure 2 a

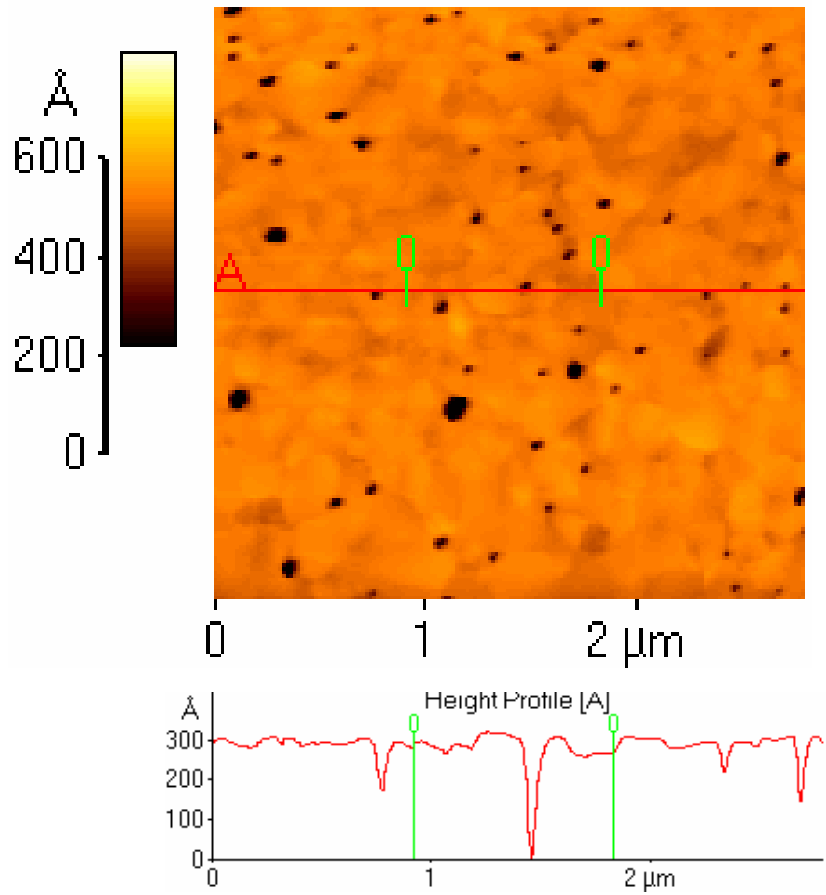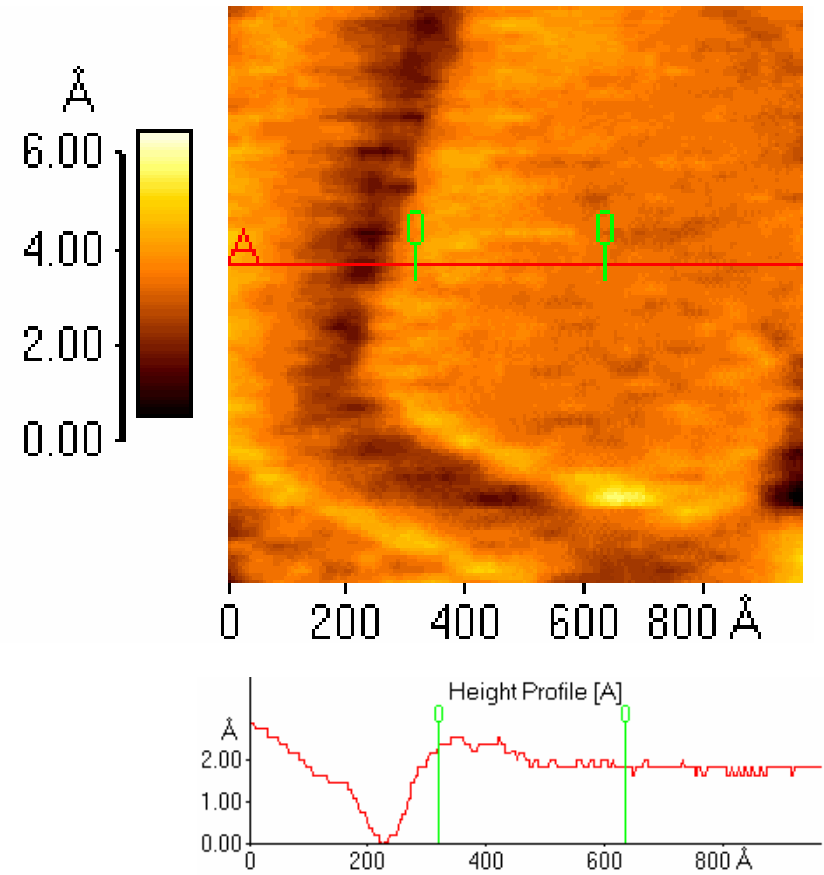

Figure 2 b

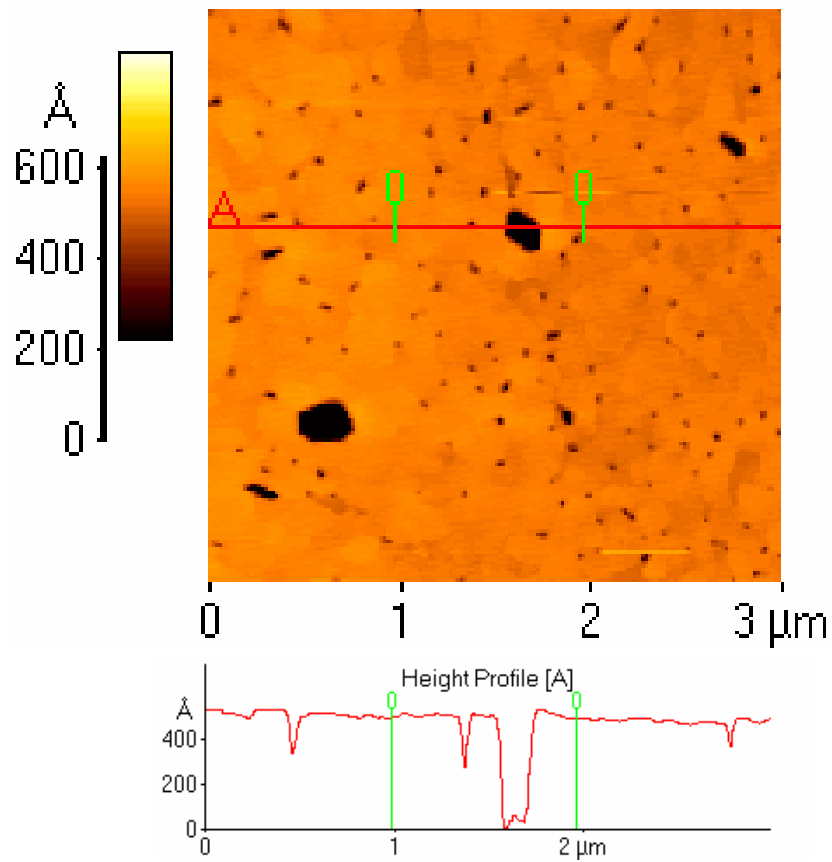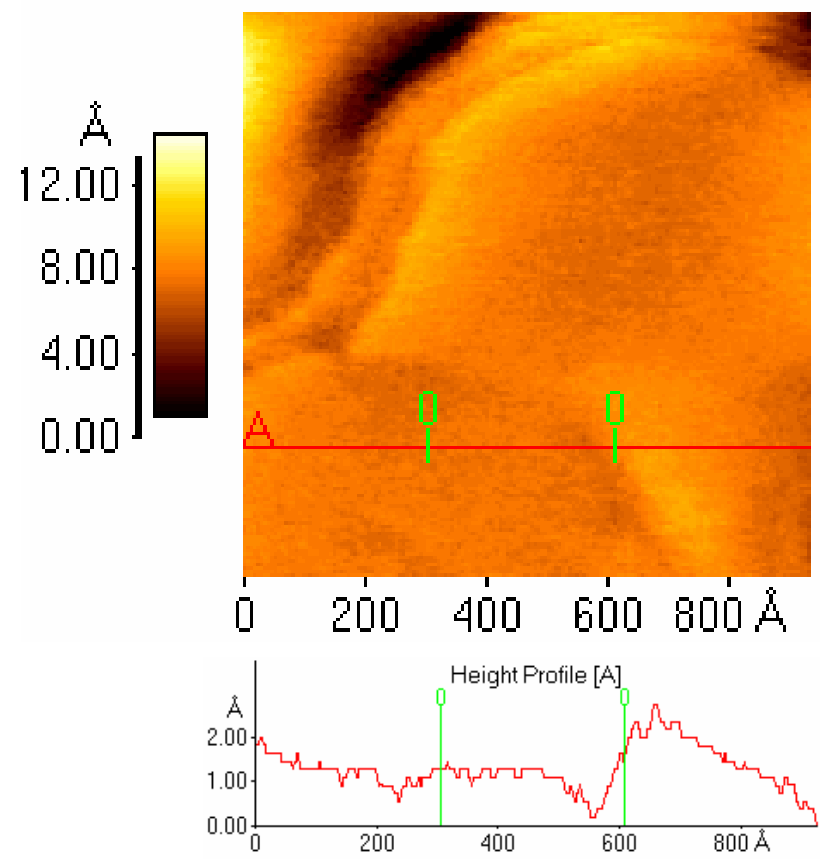

Figure 2 c

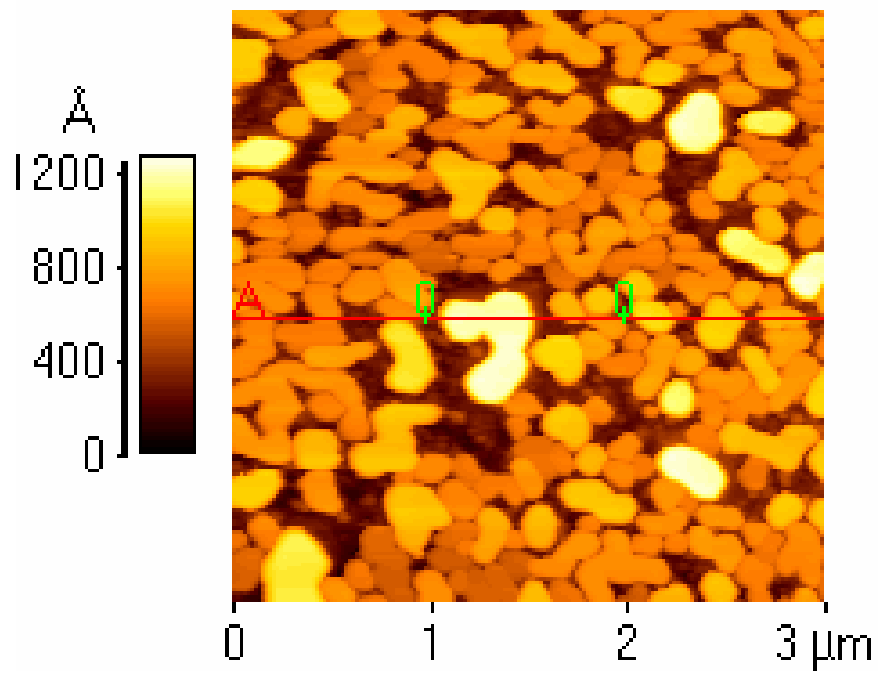
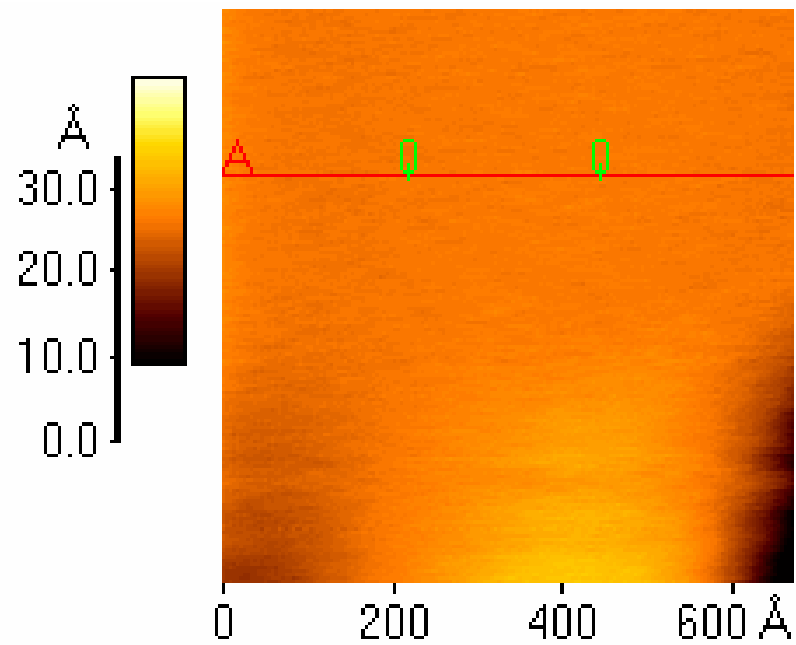
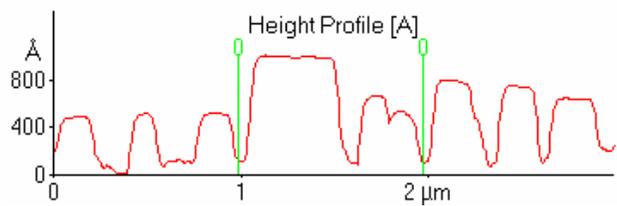
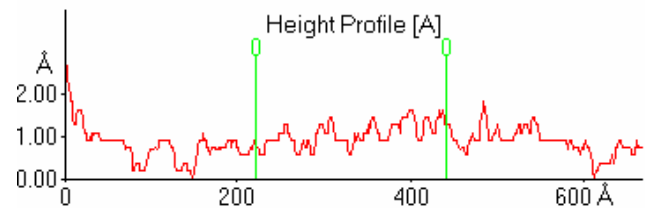

Figure 2 d

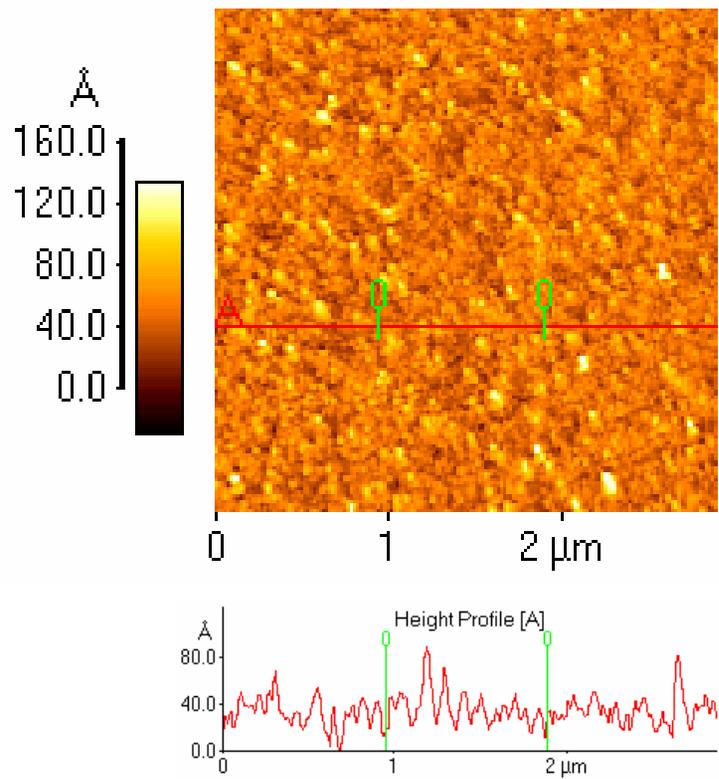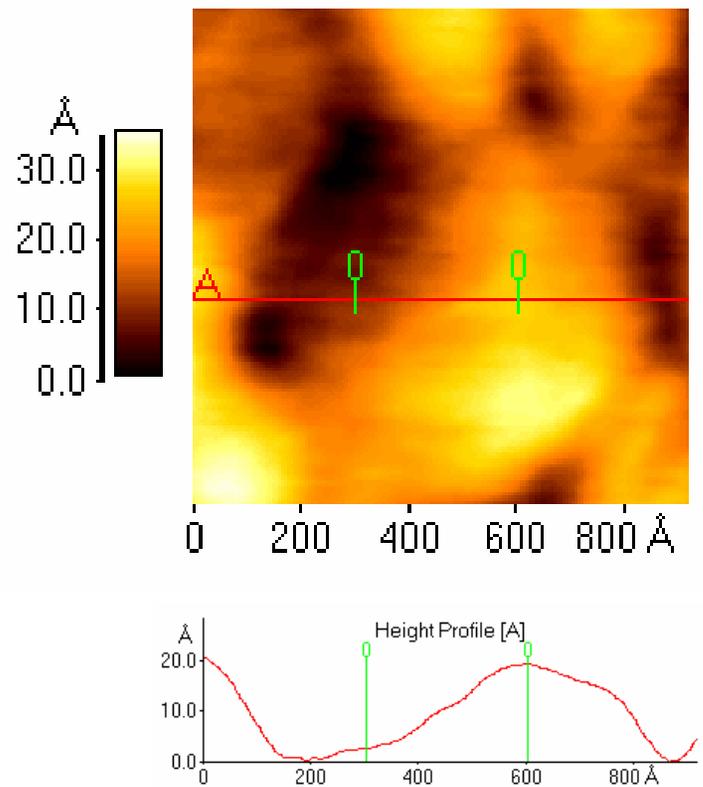

Figure 3 a

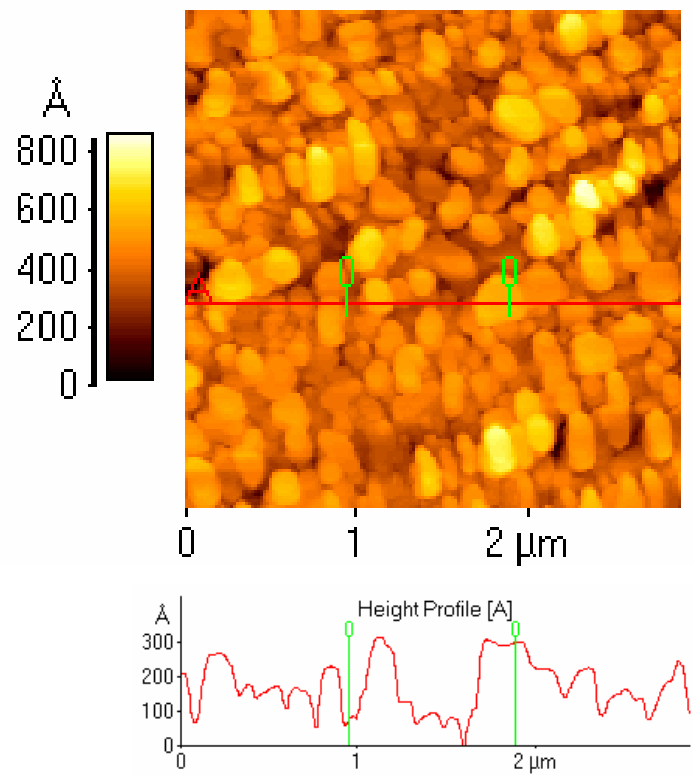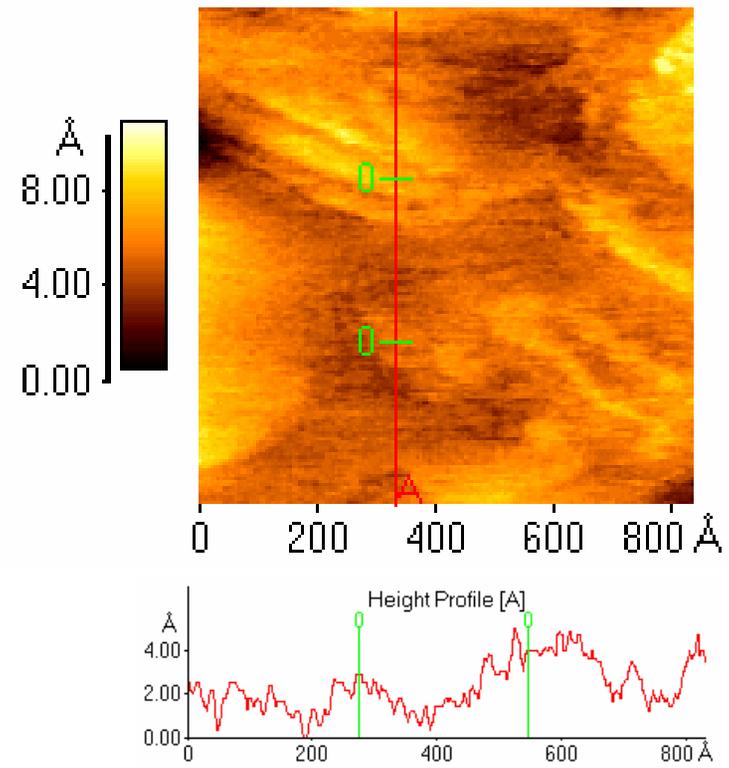

Figure 3 b

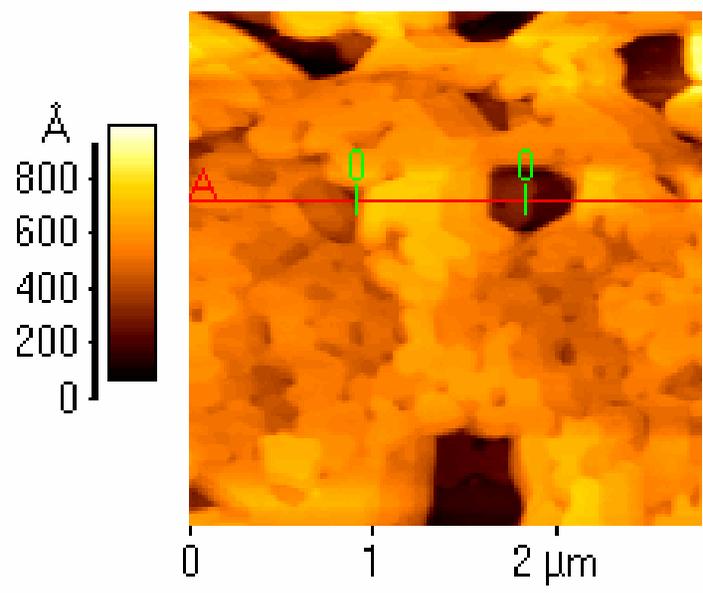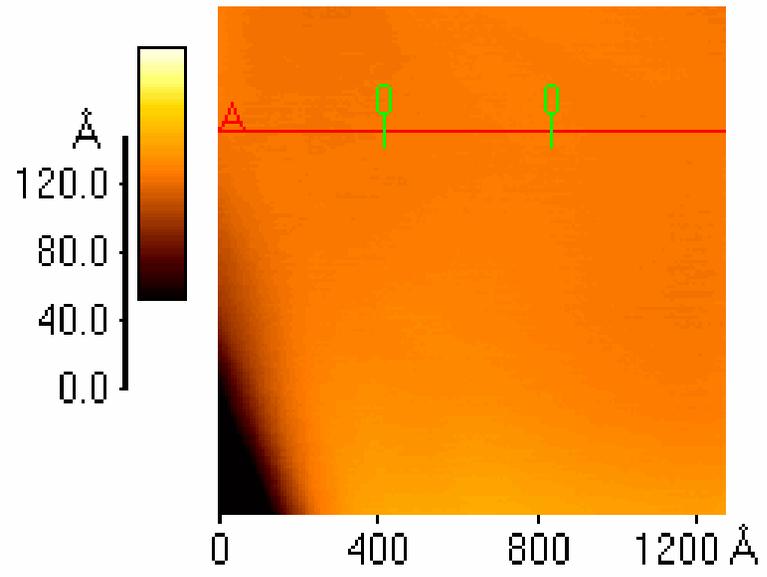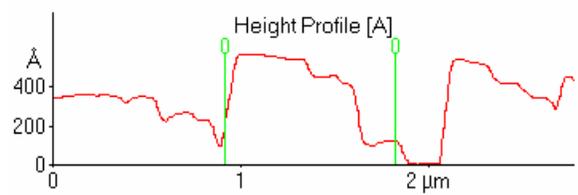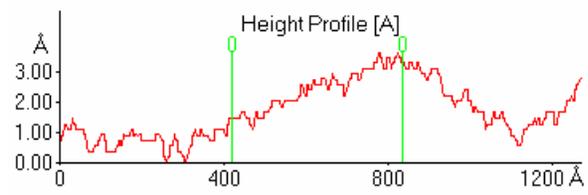

Figure 3 c

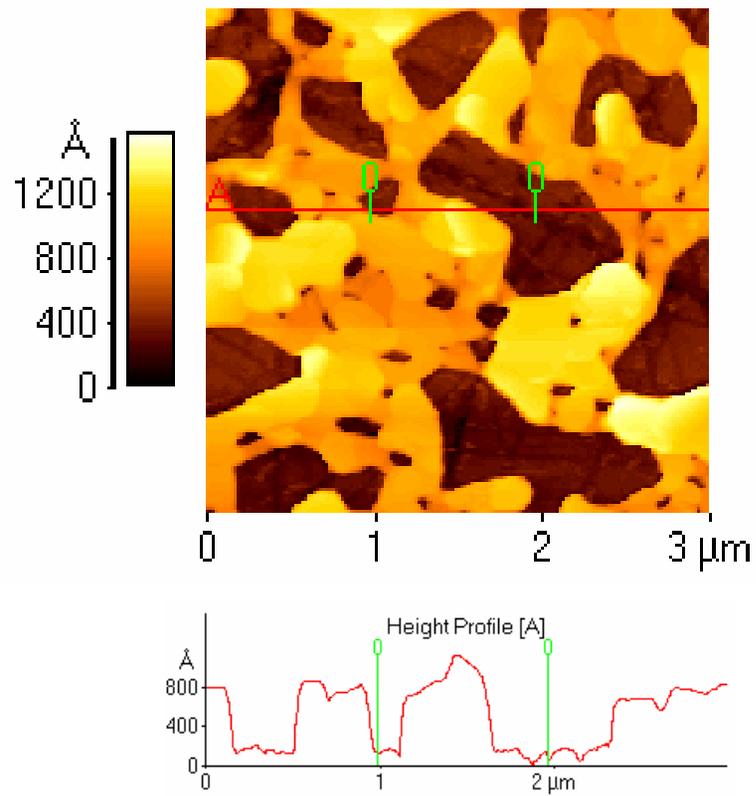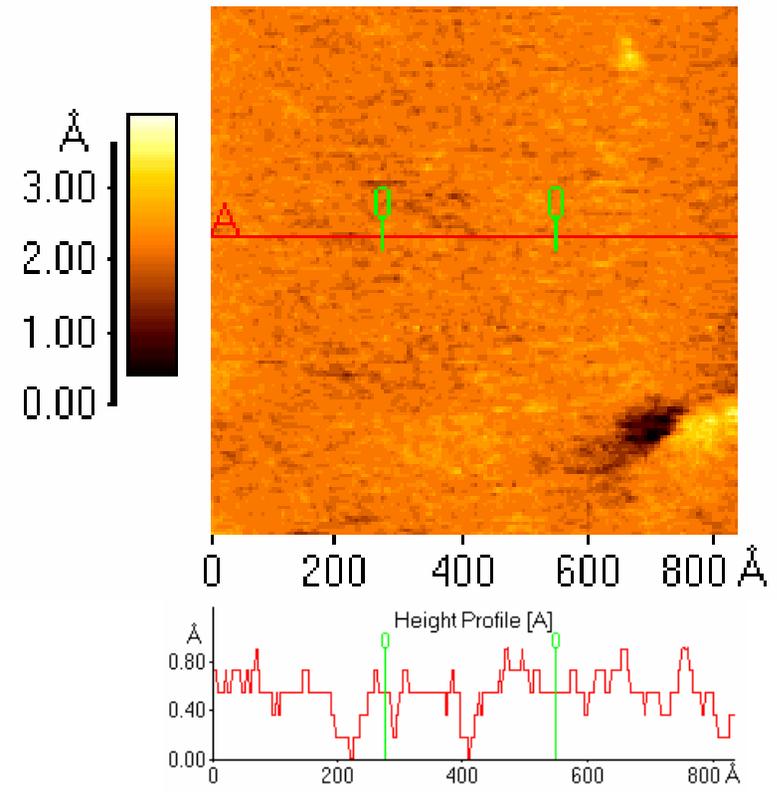

Figure 3 d

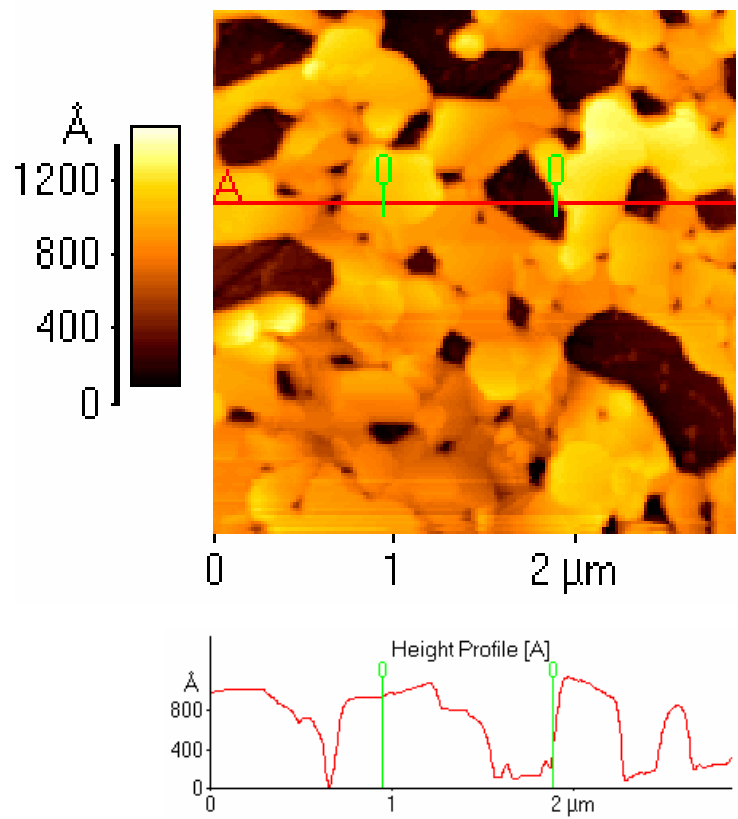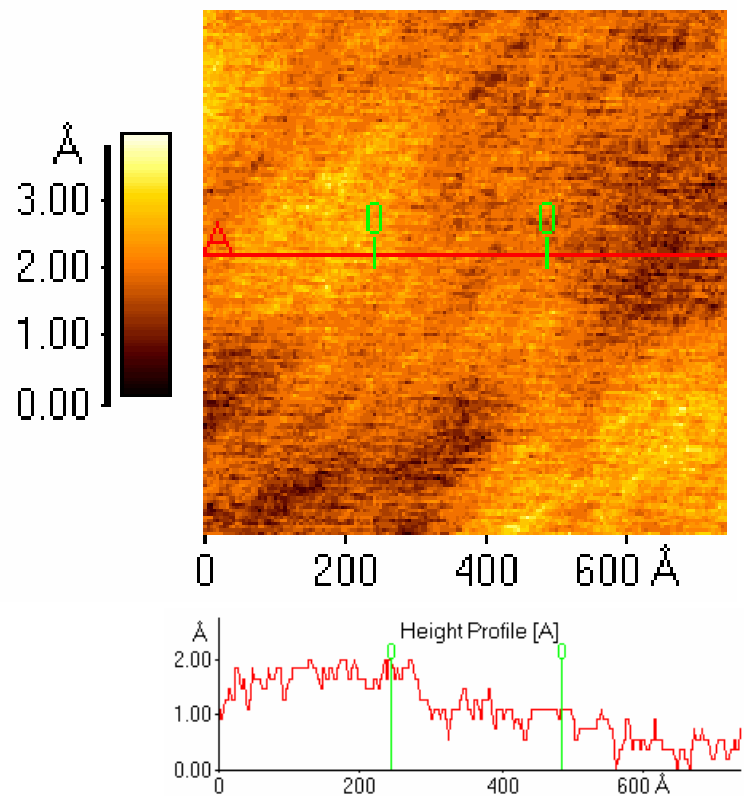

Figure 3 e

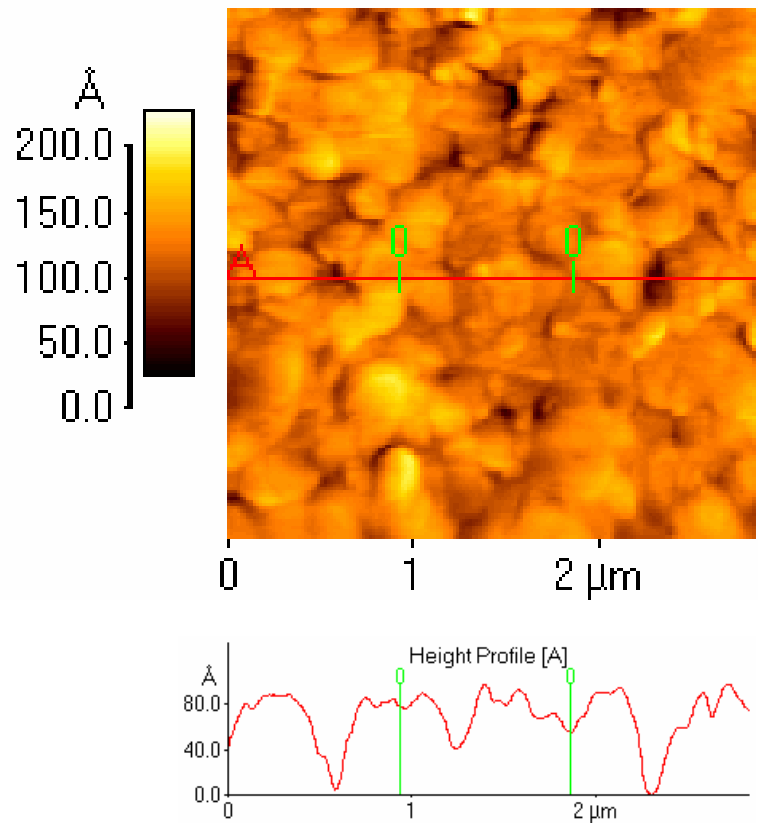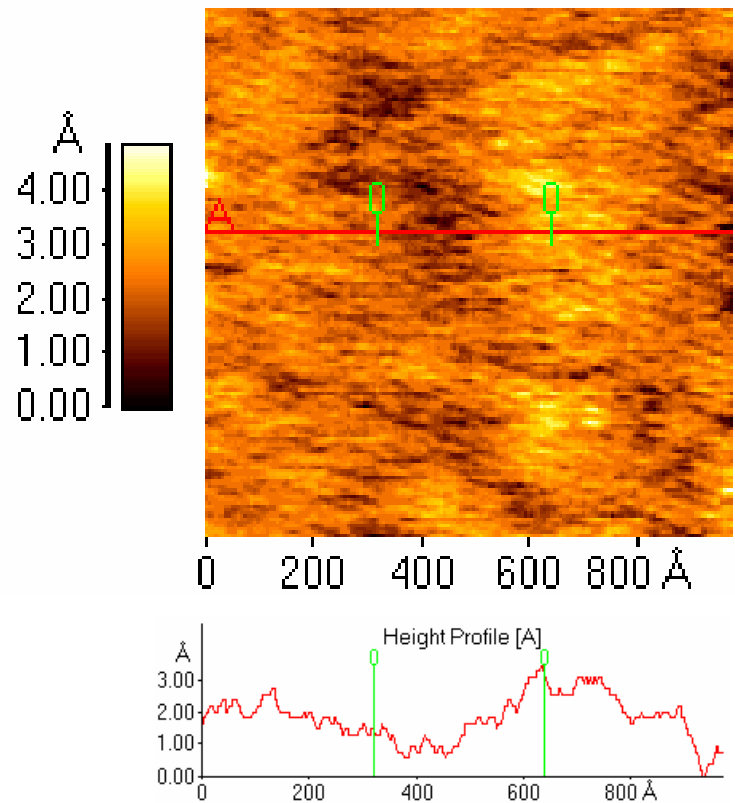

Figure 3 f

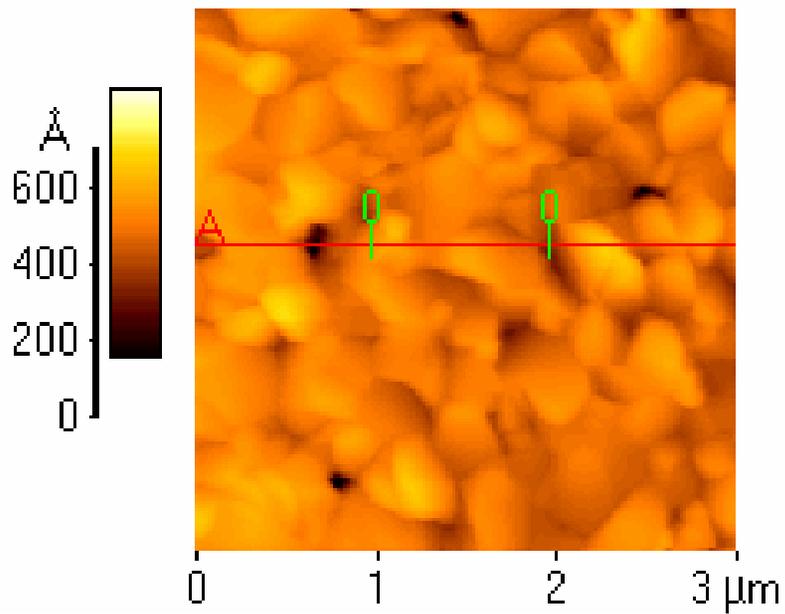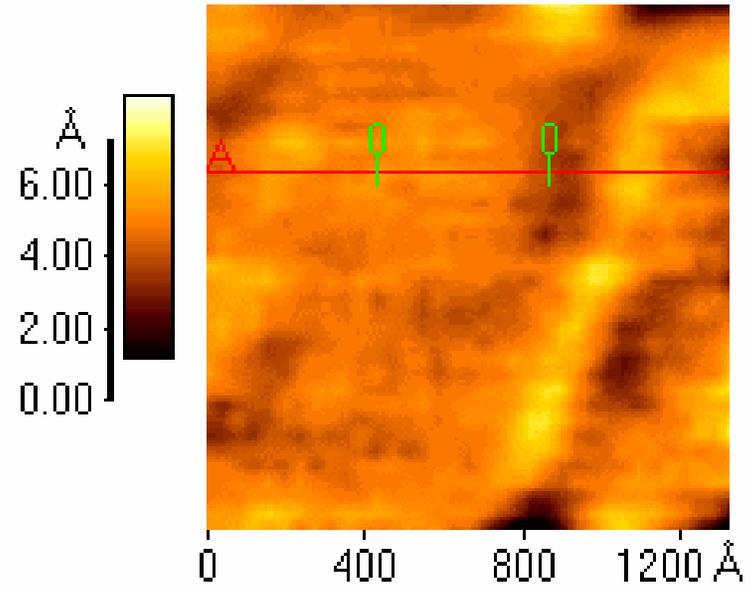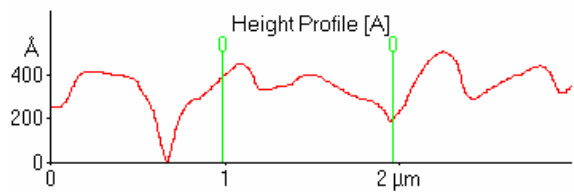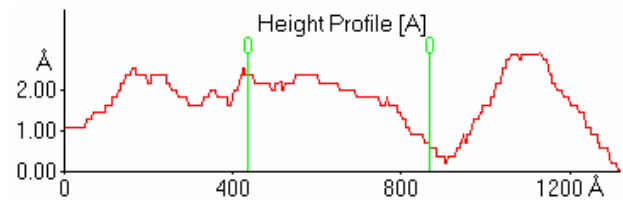

Figure 3 g

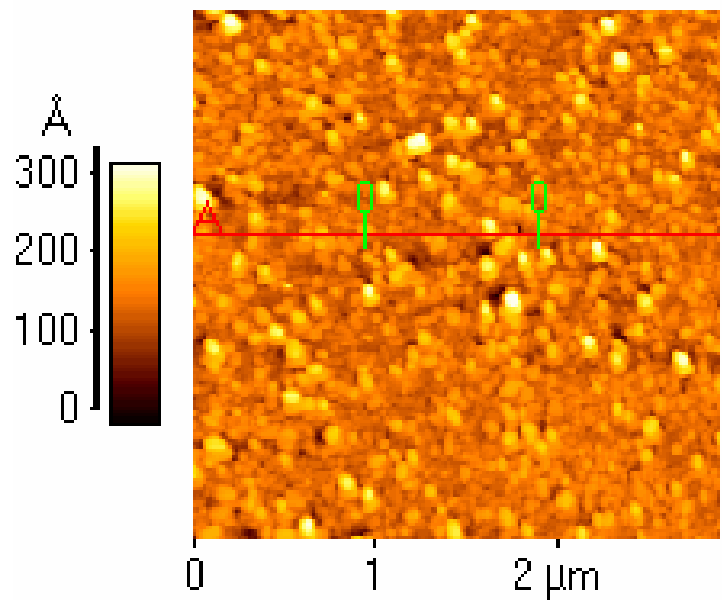
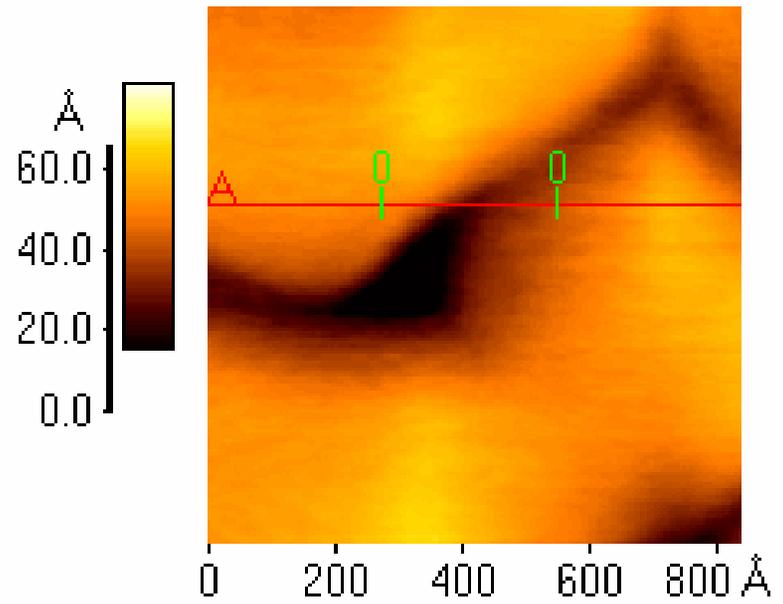
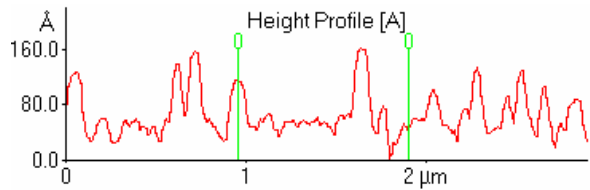
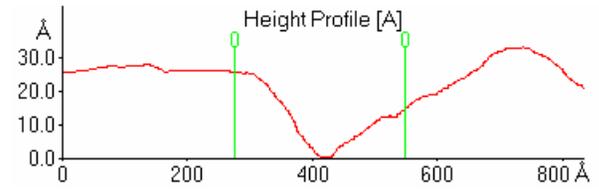

Figure 4 a

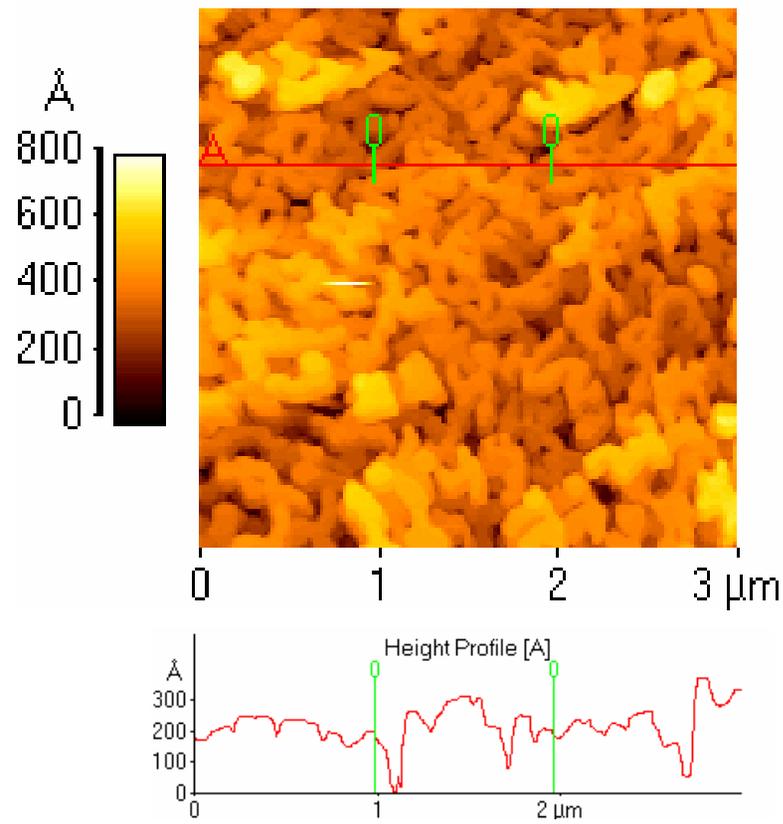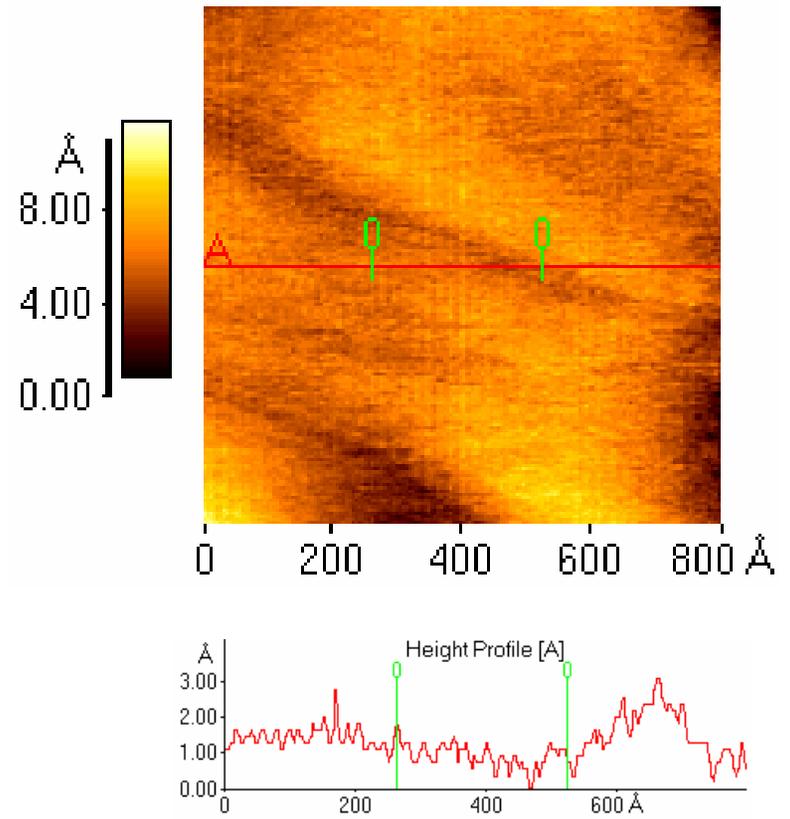

Figure 4 b

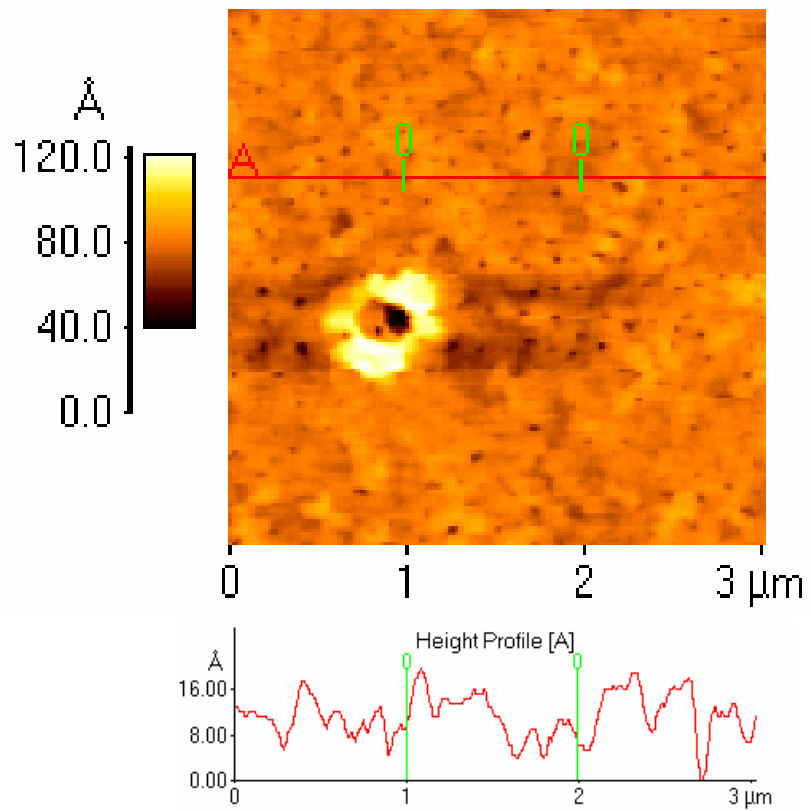

Figure 4 c

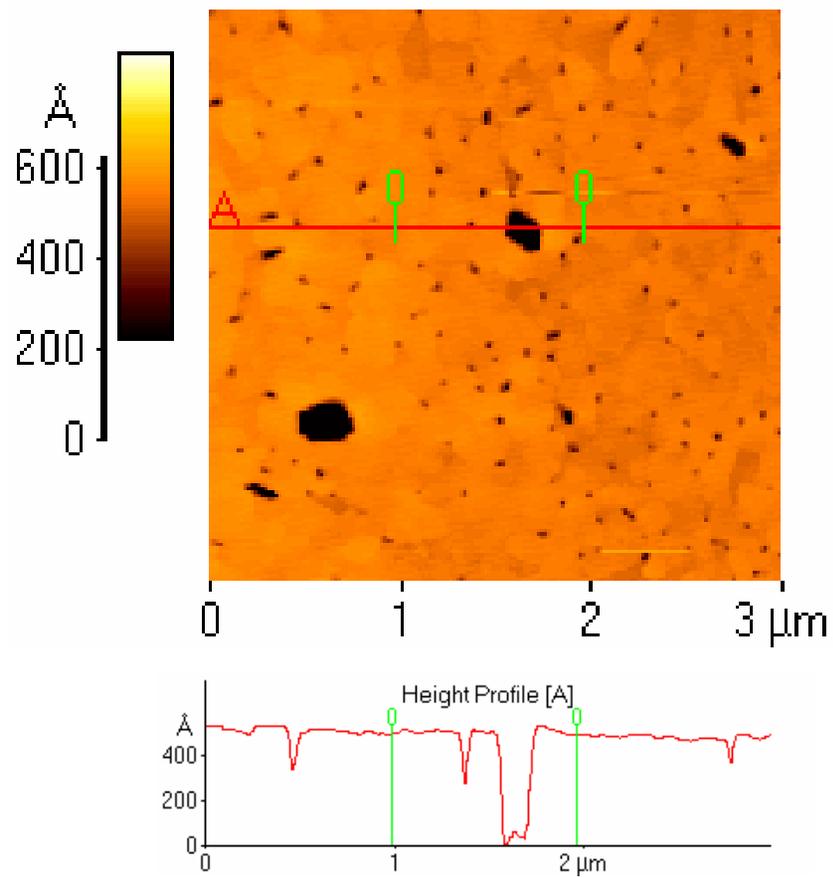
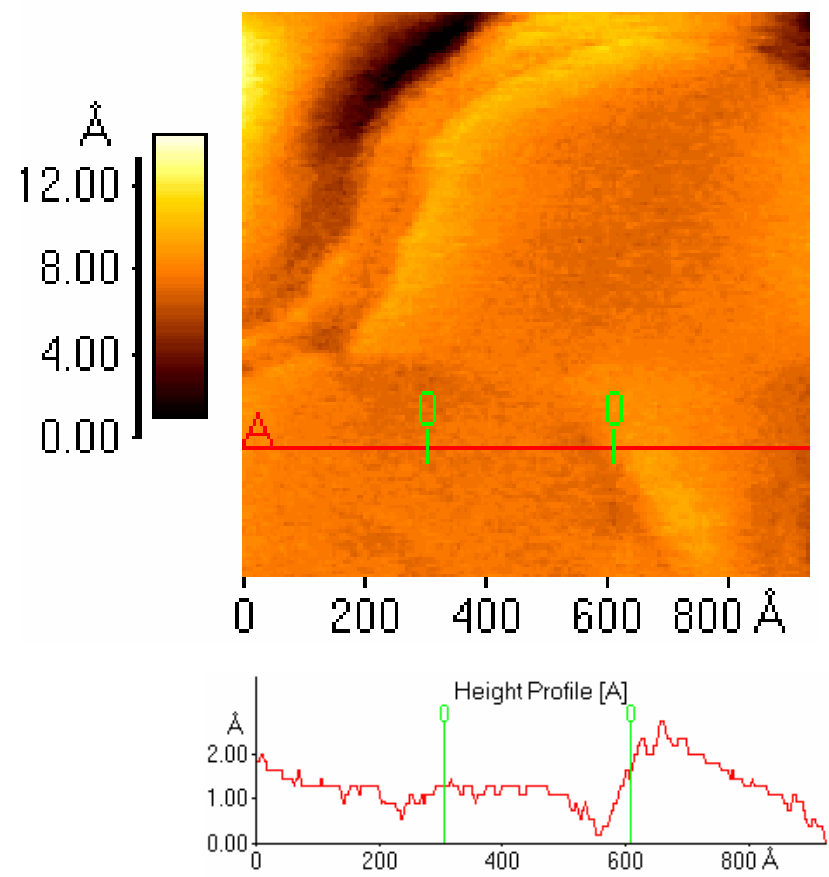

Figure 4 d

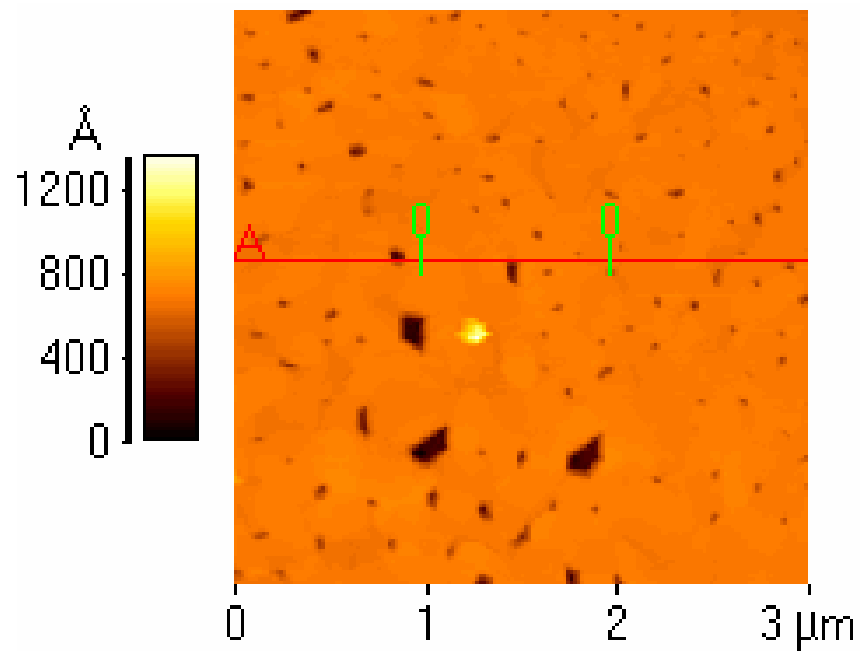
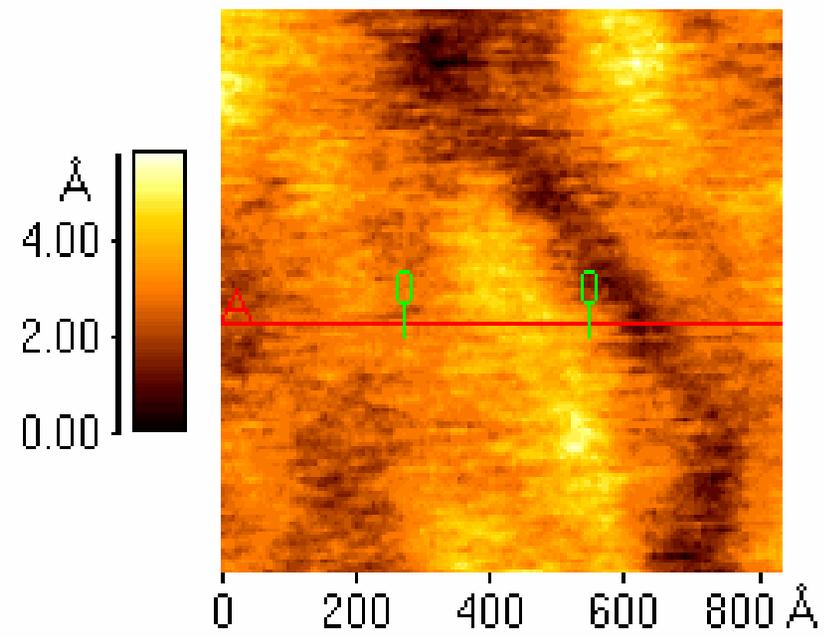
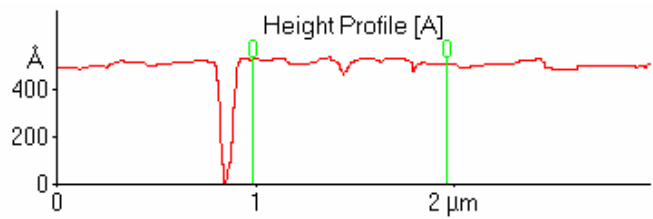
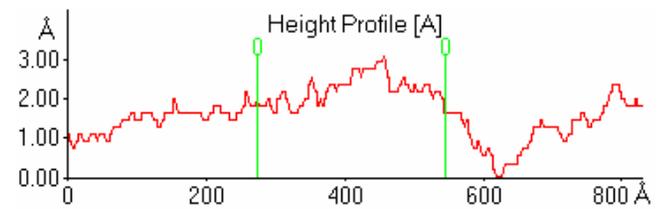

Figure 4 e

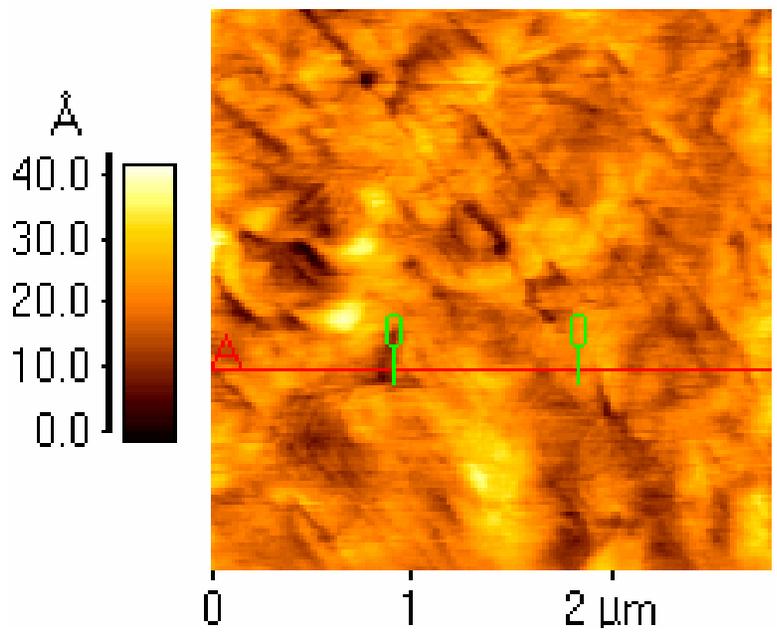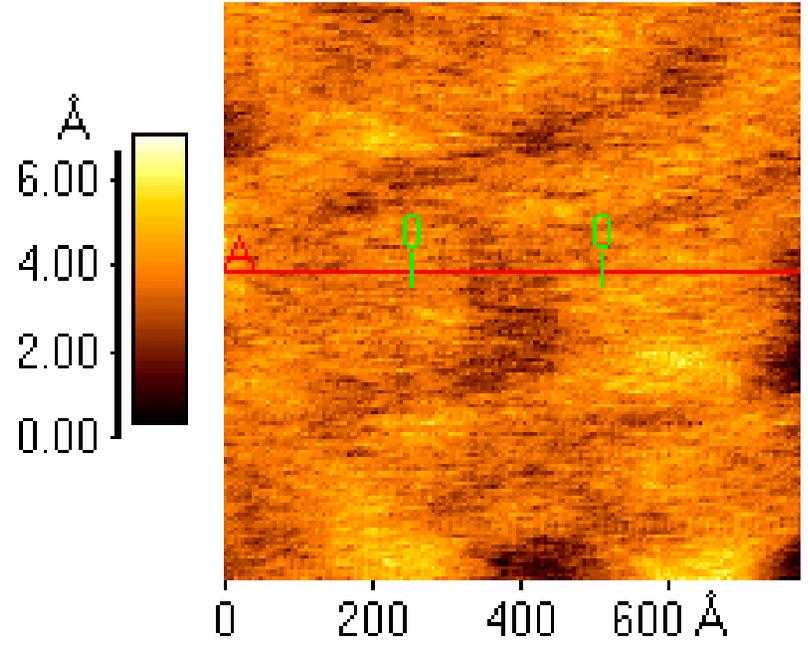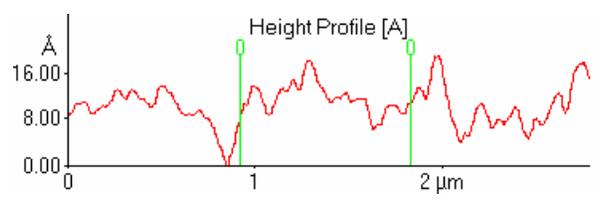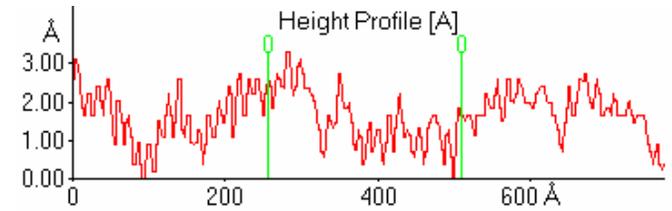

Figure 4  f

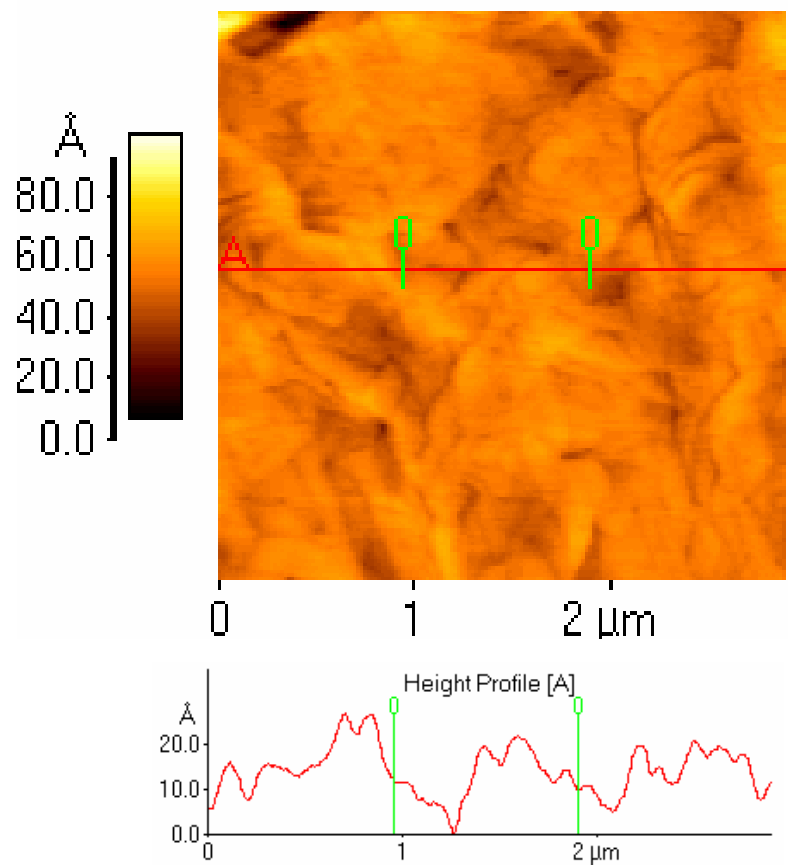 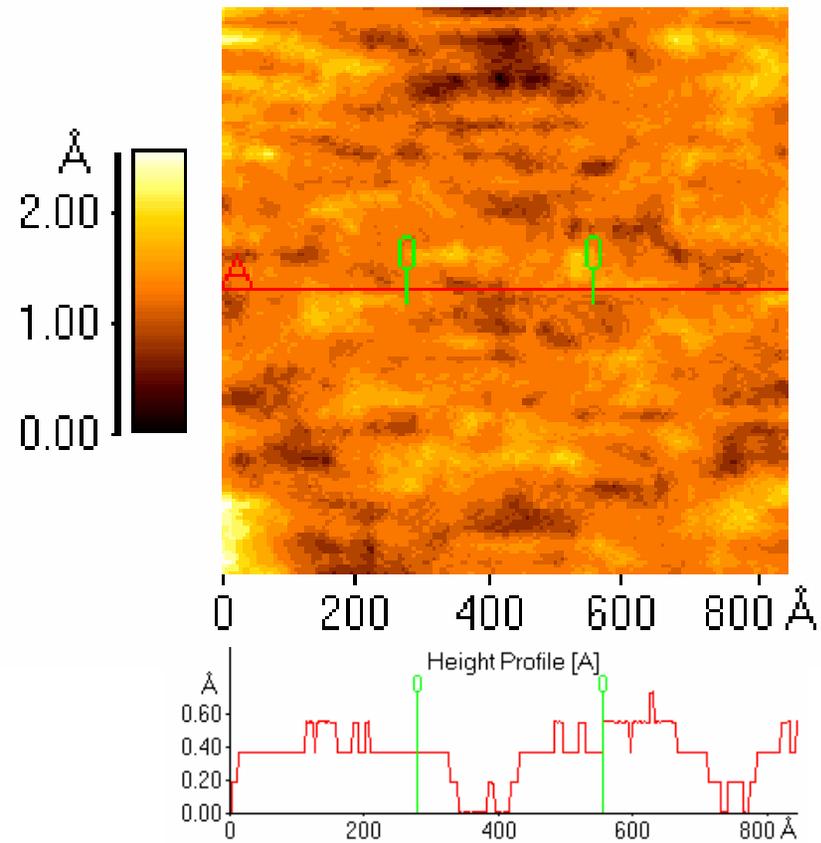

Figure 4 g

a) 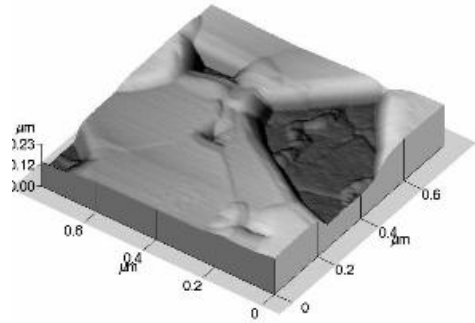

b) 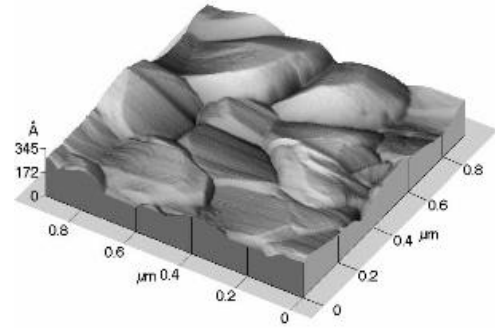

c) 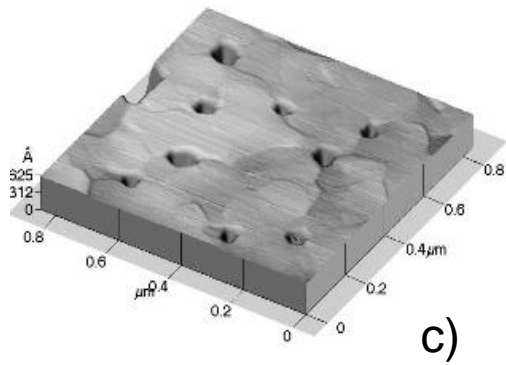

d) 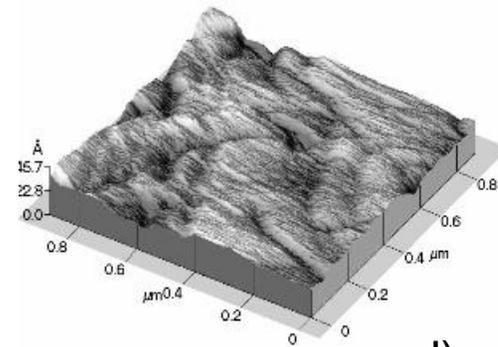

Figure 5

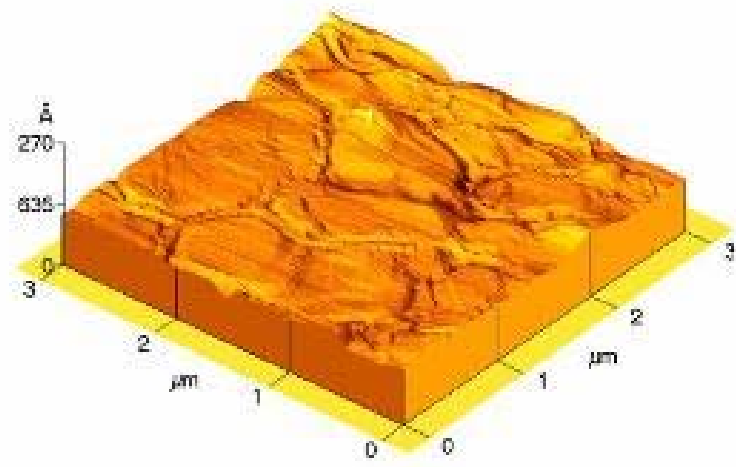

Figure 6

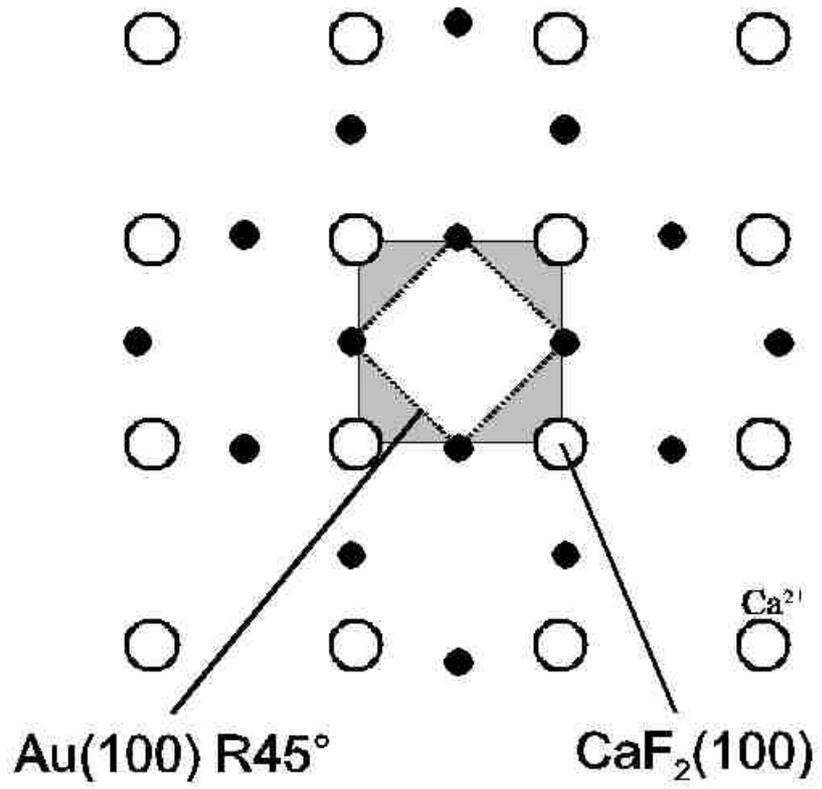

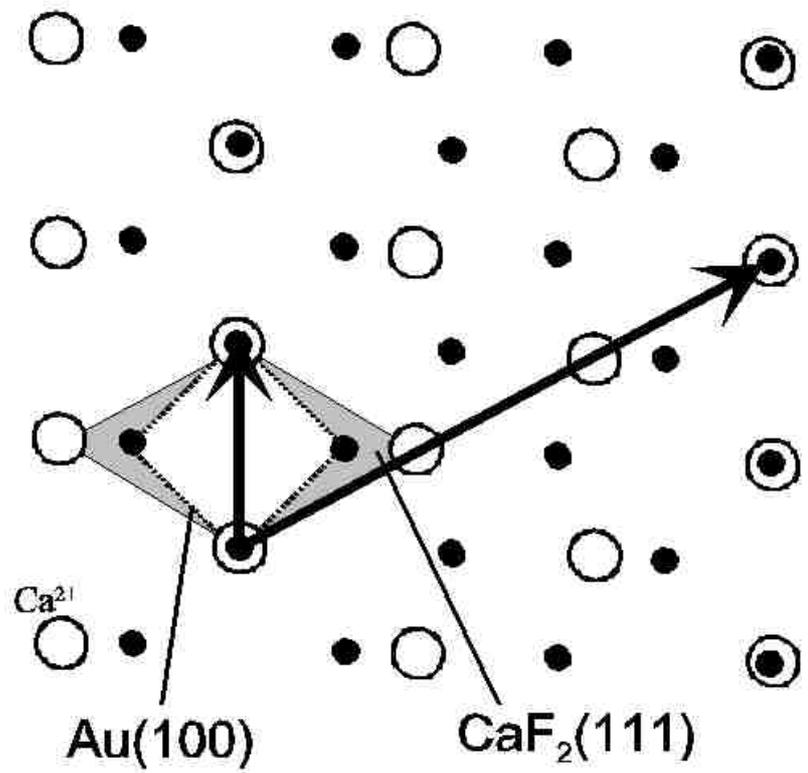

Figure 7